\documentclass[12pt]{iopart}
\usepackage{graphicx}

\newcommand{\be}{\begin{equation}}
\newcommand{\ee}{\end{equation}}
\newcommand{\bea}{\begin{eqnarray}}
\newcommand{\eea}{\end{eqnarray}}
\newcommand{\beal}{\begin{align}}
\newcommand{\eal}{\end{align}}
\newcommand{\bespl}{\begin{split}}
\newcommand{\espl}{\end{split}}

\newcommand{\nslash}{\kern 0.2 em n\kern -0.50em /}
\newcommand{\kslash}{\kern 0.2 em k\kern -0.45em /}
\newcommand{\pslash}{\kern 0.2 em p\kern -0.50em /}
\newcommand{\Sslash}{\kern 0.2 em S\kern -0.50em /}
\newcommand{\Pslash}{\kern 0.2 em P\kern -0.50em /}
\newcommand{\Rslash}{\kern 0.2 em R\kern -0.50em /}

\begin{document}

\title[Extraction of contributions to the Sivers]
{Extraction of finite-energy contributions to the Sivers asymmetry 
from the analysis of exclusive proton-proton scattering. 
} 

\author{A.~Bianconi}
\address{Dipartimento di Chimica e Fisica per l'Ingegneria e per i 
Materiali, Universit\`a di Brescia, I-25123 Brescia, Italy, and\\
Istituto Nazionale di Fisica Nucleare, Sezione di Pavia, I-27100 Pavia, 
Italy}
\ead{andrea.bianconi@bs.infn.it}

\begin{abstract}
I study the effect of scalar and spin-orbit  
rescattering terms, in the 
production of a nonzero Sivers-like asymmetry 
in proton-proton collisions (inclusive production 
and Drell-Yan) at moderate 
center of mass energies $\sqrt{s}$ $<$ 15 GeV 
and transverse momentum up to 3 GeV/c.  
An ultrarelativistic generalization 
of the Glauber formalism is here used to (i) fit the 
scalar and spin-orbit interaction terms on proton-proton 
elastic scattering data, including analyzing power, (ii) 
transfer such information to inclusive proton-proton 
scattering. It is shown that the phenomenological interactions 
responsible for the nonzero analyzing power in proton-proton 
elastic scattering produce a relevant nonzero 
analyzing power in inclusive processes associated 
with proton-proton collisions. This could represent a 
relevant (possibly higher-twist) contribution to the Sivers 
asymmetry. 
\end{abstract} 


\maketitle

\section{Introduction}

\subsection{General background}

The problem of the study and measurement of T-odd distributions 
in hadron-hadron 
scattering has recently acquired 
a certain relevance and quite a few related experiments 
have been thought or scheduled for the next ten 
years\cite{panda,assia,pax,rhic2,compassDY}. 

In particular several studies and models have been proposed 
for the Sivers distribution function\cite{Sivers}. Its possible 
existence as a leading-twist distribution was 
demonstrated\cite{BrodskyHwangSchmidt02,Collins02,
JiYuan02,BelitskyJiYuan03} recently, and 
related\cite{JQVY06} to previously 
studied T-odd mechanisms\cite{EfremovTeryaev82,QiuSterman91}. Some 
phenomenological forms for its dependence on $x$ and $k_T$ have been 
extracted\cite{Torino05,VogelsangYuan05,CollinsGoeke05,BR06a,Anselmino08} 
from available data\cite{Star,Hermes,Compass,Phenix,Transversity2008}. 

While studies of general 
properties\cite{Pobylitsa03,DalesioMurgia04,Burkardt04,
Drago05,BR_JPG,GoekeMeissnerMetzSchlegel06,BoerVogelsang06,
BR_JPG2,Entropy3,Koike07,BDGMMS07,Bacchetta08} 
of T-odd functions 
relate these functions with a wide spectrum of phenomena,  
quantitative models mostly follow the general scheme suggested 
in \cite{BrodskyHwangSchmidt02}. A known quark-diquark 
spectator model\cite{JakobMuldersRodriguez97} is extended by 
including single boson  
exchange\cite{BoerBrodskyHwang03,GambergGoldsteinOganessyan03,
BacchettaSchaeferYang04,LuMa04}. 
In the case of \cite{Yuan03} the 
unperturbed starting model was a Bag model, and for \cite{Courtoy08}
a constituent quark model. 

Here, I want to follow a different approach, that may be of 
interest in the case of intermediate energy measurements, and 
that relates Single Spin Asymmetries 
(SSA) in inclusive processes with SSA in exclusive 
processes. 

A well measured SSA in an exclusive 
channel is the 
normal vector analyzing power  
measured in elastic proton-proton 
scattering at energy 20-30 GeV and $Q_T$ $=$ 1$-$3 GeV/c 
(see \cite{LLL93,
Crabb90,Cameron85,Peaslee83,Hansen83,
Antille81}). 

In the energy range 20-28 GeV, and 
for $Q_T-$ near 3 GeV/c, this observable is unexpectedly 
large  
and roughly energy-independent (see fig.8 later in this work).  
At energies over 30 GeV or $Q_T$ over 3 GeV/c 
it is not measured. 

A nonzero analyzing power 
in $pp$ $\rightarrow$ $pp$ 
requires interference between helicity-flip and helicity 
non-flip amplitudes with 90$^o$ phase 
difference (see e.g. \cite{GoldsteinMoravcsik82} 
or \cite{LLL93} for reviews 
on this and related 
points). 
For this reason, 
in PQCD helicity-conserving processes this analyzing power must 
be very small, and 
the origin of the phenomenon 
has not yet a commonly accepted explanation.  
Attempts to explain it may be traced back to 
ref.\cite{LandshoffPolkinghorne71} in a non-QCD context. 
For the region of semihard $Q_T$ $=$ 1$-$3 GeV/c 
many nontrivial models have been 
proposed\cite{GST79,Farrar86,BCL79,BSW79,
Anselmino82,BrodskyTeramond,
GoldsteinMoravcsik85,KazacsTucker88,
Lipkin,Hendry,RalstonPire86,Tomozawa,Wolters,
DurandHalzen,SakamotoWakaizumi,ACM81,Bourely89}, but 
a ``standard'' 
model, in the sense of a commonly accepted model for this  
effect, does not exist at present. It must be observed that 
for some of the previous models (e.g.\cite{RalstonPire86}) 
the effect should persist at much larger energies 
than 30 GeV, 
while other ones select a preferential energy range, over 
which the effect should be suppressed. 

The idea underlying the present work is that the interactions 
producing 
such an asymmetry are also active in inclusive processes 
originated by hadron-hadron collisions, at the same energy come 
and transferred momenta. In such a case, they could contribute 
to a nonzero asymmetry of Sivers-like kind. 
So a scheme can be imagined, 
allowing for information transfer from elastic proton-proton scattering 
to proton-proton 
inclusive processes. This is attempted in the following. 

I do not propose nor adopt any model for the proton-proton elastic 
scattering amplitude. But I postulate a general form for 
a set of amplitudes contributing to this process at quark 
level. These amplitudes become initial state interactions 
in inclusive processes, where their presence admits for 
a nonzero Sivers asymmetry. 

\subsection{This work}

The class of processes I want to consider here is the one of 
SSA in inclusive 
collisions between an unpolarized 
proton and a transversely polarized proton. In particular, 
azimuthal asymmetries in inclusive 
production of mesons or (virtual/real) photons. 
For fixed target experiments, the kinematics of 
interest is the one with moderate beam energy 10-60 GeV, and  
transverse momentum in the semihard regime 1-3 GeV/c. 

To find nonzero T-odd quantities like the Sivers 
function one needs rescattering interactions 
between the active quarks and the surrounding particles. 
In particular, imagining that a hadron-hadron 
inclusive hard process 
is triggered by a hard collision between 
a quark in the projectile hadron (active quark) and a parton 
in the target hadron, this work is centered on the rescattering 
interactions between the active quark and the target hadron 
(initial state interactions, ISI). 

First, I introduce 
a very simple bound state in light-cone coordinate 
representation 
for the active quark in the projectile proton. For this state, 
I speak of ``unperturbed'' bound state, where ``unperturbed'' 
means that 
it is not affected by ISI with the target hadron. 
This state will be a starting 
point for reproducing both inclusive and exclusive processes. 
It is a two-component state (each component representing 
positive or negative transverse quark spin). In Appendix A it 
is shown that in ultrarelativistic conditions these two components 
contain all the relevant independent information on the quark 
state. 

Before the hard scattering event, this state is affected 
by ISI. 
The precise form and the parameter values for ISI 
are extracted from the phenomenology of elastic proton-proton 
scattering in semihard conditions. 
I assume that the interactions producing a nonzero analyzing 
power in proton-proton elastic scattering at the required 
kinematics 
may be rewritten in terms of interactions between a 
projectile quark and a target hadron, where the former is 
bound to a projectile hadron, and the 
latter has a continuous structure. 

The scheme used for this aim 
is an ultrarelativistic generalization 
of the Glauber method\cite{Glauber59}. 
Starting from the fact that the fitted data include unpolarized 
scattering and single normal spin analyzing power, 
consideration of the number of constrained amplitudes 
reduces the considered ISI to a sum of scalar 
and spin-orbit interaction terms, in a 2x2 formalism. 

Fitting parameters on proton-proton data does not allow for 
a strict flavor separation. However, it implies 
u-quark dominance.  So in the following the 
expression ``quark interactions'' 
mainly means ``u-quark interactions''.

In Section II, some general definitions are presented and 
discussed, in particular the definitions of unpolarized 
quark distribution and Sivers-like 
analyzing power, in terms of empiric variables on the one side and 
of quark operators on the other side. 
Also, the quark 
unperturbed bound state in coordinate representation 
is introduced. Some related details are put into 
Appendix A section. 

In Section III, an operator describing ISI  
is introduced. Having to modify 
the above two-component state, this operator 
consists of a 2x2 matrix operator, that is written 
in eikonal form as the exponential of  
a set of scalar plus spin-orbit matrices. 

I describe and discuss the theoretical foundations and several 
details of the method allowing me to extract the form and 
parameters of the ISI operator 
from elastic proton-proton scattering 
and to apply it to the calculation of inclusive quark 
distributions.  Some details are put into Appendix B Section. 

In Section IV the parameters of the ISI operator are fitted to 
reproduce data on proton-proton elastic scattering at 
beam energy 20$-$50 GeV and transferred momentum 1$-$3 GeV/c, 
and MRST\cite{MRST} 
unpolarized u-quark collinear distribution at 
$Q^2$ $=$ 16 GeV$^2$. 

In Section V the rescattering operator is used to calculate 
the unpolarized quark distribution, including 
Sivers-like asymmetry, for $x$ in the valence region and transverse 
momenta up to 3 GeV/c. 

It must be remarked that this work is subject to some limits: 

1) Computational limits. I face stability problems  
in calculating Fourier transforms for transverse 
momenta over 3 GeV/c (elastic scattering) or 2.7 GeV/c (inclusive 
at $x$ $=$ 0.3). These problems get worse at 
increasing x, so my analysis centers  
at $x$ $=$ 0.3. This value guarantees that we are in the 
valence region, and that numerical results are reliable. 

2) Limitedness of the data set used for fixing the parameter 
values. 
Proton-proton scattering is not the only exclusive 
process from which information on rescattering in hadron-hadron 
hard processes may be extracted, although it represents the 
most complete and precise set of available data 
in this respect. 

3) Twist identification. 
The formal apparatus introduced here 
implies leading twist results at the condition that the 
key parameters of the interaction operator become 
energy-independent at large energies. Since there is presently no way 
to decide how these parameters behave asymptotically, 
it is not possible to establish whether the found effect is  
a leading or a higher twist one. 
If it is a relevant higher twist, 
it should become negligible (compared to leading twist effects) 
at energies like 100 GeV. Exploiting the fact that the 
measured analyzing power is energy-independent from 20 to 
30 GeV, I assume that the considered 
interactions are relevant in the energy range 10-60 GeV. 
For what happens over this range, I cannot formulate hypotheses. 

Here the terms ``Sivers asymmetry'' and ``Sivers effect'' 
are preferred to 
``Sivers function''. The last one is appropriate in the case of 
a leading twist contribution. To conform with common 
notation, in the result section I will name ``Sivers function'' 
a quantity that may be extracted from single spin asymmetries  
in inclusive processes. This 
quantity must be meant as a measured function, whose 
theoretical interpretation may be and may be not the one of 
a Sivers function. 

As a last remark, data and distributions shown in figs. 6, 7, and 8, 
together with the original references, have been 
reconstructed thanks to the Durham HEP database\cite{HEPDATA}.  

\section{The general scheme - no rescattering.} 

\subsection{Basic variables.}

Apart for some points where it is clearly 
specified, all variables will refer 
to the center of mass of the colliding hadrons. Let 
$\vec b$ $=$ $(b_x,b_y)$ 
be the quark impact parameter and 
$\vec k_T$ $=$ $(k_x,k_y)$  
the transverse momentum conjugated 
with it. Let $P_+$ be the large light-cone component of the 
hadron momentum, so that $xP_+$ is the quark $(+)$ momentum 
conjugated with $z_-$. Assuming by default the validity of a 
standard factorization scheme\cite{CollinsSoperSterman,Bodwin}, 
the fourth coordinate $z_+$ plays no role and is fixed to zero. 

I substitute $z_-$ with the rescaled coordinate 
\begin{equation}
\xi\ \equiv\ P_+z,\hspace{0.5truecm} 
\rightarrow \hspace{0.5truecm}
P_+ \int dz_- exp(-ixP_+z_-)\ =\ 
\int d\xi exp(-ix\xi) 
\label{eq:fourier0}
\end{equation}
not to work with a singularity of the 
Fourier transform 
in the infinite momentum limit $P_+$ $\rightarrow$ $\infty$. 

Contrary to the ordinary treatment of the problem, where one 
works on a two-point correlation operator deriving from 
a set of squared one-point amplitudes,  
I develop most of the work at 
the level of one-point amplitudes, 
square them and then sum over 
the relevant states. 

Since the inclusive process is described here in terms of squared 
amplitudes, and these amplitudes are calculated before being 
squared, $\xi$ is not bound 
to be positive, as it happens in the ordinary treatment based on a 
two-point correlator with intermediate real states. In that 
case $\xi$ has the meaning of the difference between 
the light-cone positions of two points. 
In this work it describes the light-cone position of one of the two only. 

\subsection{Two-component transverse spin formalism.}

I consider a quark inside a hadron with a given 
spin projection $S_y$ $=$ $+1/2$. The quark is supposed to 
present a nonzero transverse momentum along the $\hat x$ direction, 
and is described by a two-component quark 
spinor 
\begin{equation}
\vec \psi\ \equiv\ (\psi_+,\psi_-). 
\label{eq:two_spinor}
\end{equation}

These two components are the components of the full 4-spinor 
$\Psi$ describing a free quark on 
two 4-spinors $\Psi_{T+}$, $\Psi_{T-}$:  
\begin{equation}
\Psi\ \equiv\ \psi_+\Psi_{T+}\ +\ \psi_-\Psi_{T-}
\label{eq:states1}
\end{equation}
where 
\begin{equation}
\Psi_{T+}\ \approx\ (\vec\phi_+, \vec\phi_-),\ 
\Psi_{T-}\ \approx\ (\vec\phi_-,\vec\phi_+)
\label{eq:states2}
\end{equation}
and $\vec\phi_\pm$ are 2-component eigenstates of 
\begin{equation}
\hat \sigma_y \vec \phi_\pm\ =\ \pm \phi_\pm. 
\end{equation}
where the scattering plane is formed by the directions ``z'' 
(longitudinal) and ``x'' (transverse).  

In the Appendix A section it is demonstrated (i) that in the 
ultrarelativistic limit the above transverse spinors give 
a full description of the state of a free quark, (ii) that 
\noindent
\begin{equation}
\bar{\Psi} \gamma_+\Psi\ \approx\ |\psi_+|^2\ +\ |\psi_-|^2,
\label{eq:trace1}
\end{equation}
that becomes an equality in the u.r. limit. 
The $\gamma_+-$projection  
produces the 
distribution functions associated to an unpolarized 
quark with positive $z-$direction in an infinite momentum frame.

\subsection{TMD quark distribution and Sivers-like analyzing power}

I consider the transverse momentum dependent (TMD) quark 
distribution $q(x,\vec k_T)$
of unpolarized quarks in a hadron with 
$y-$polarization $+1/2$. 
This may be defined in a phenomenological and in a theoretical way. 

The phenomenological definition adopted here is in agreement with 
the so-called ``Trento convention'' \cite{Trento}, in which 
the unpolarized quark transverse momentum dependence 
distribution has the form
\begin{equation}
q(x,\vec k_T)\ 
\equiv\ q_U(x,k_T)\ -\ 2{{\vec S\wedge \vec k_T \cdot \hat z}\over M} q_S(x,k_T)
=\ q_U(x,k_T)\ \pm\ {k_x \over M} q_S(x,k_T)
\label{eq:sivers}
\end{equation} 
\noindent 
for an unpolarized quark in a hadron with 
$\vec S$ $=$ $\pm\hat y$, moving along the $z$ direction.  

If the 
second term is scale-independent, $q_S$ is 
the Sivers function. 
In the following 
I will speak of ``Sivers asymmetry'' meaning 
$q_S k_x/2Mq_U$. 

This asymmetry can of course be isolated by 
calculating the ratio $[q_+(k_x)-q_-(k_x)]/[q_+(k_x)+q_-(k_x)]$ 
corresponding to opposite 
proton polarizations. Alternatively, one may use the difference 
$q_+(+k_x)$ $-$ $q_+(-k_x)$ for fixed proton polarization.  
The result is the same. 

So, the center of this work will be the quark distribution 
asymmetry 
$A(x,k_x)$, defined as 

\begin{equation}
A(x,k_x)\ \equiv\ 
\Bigg({{q(x,k_x) - q(x,-k_x)} 
\over {q(x,k_x) + q(x,-k_x)}}\Bigg)_{k_y = 0} 
\label{eq:ap1}
\end{equation} 

\noindent
In the following, this inclusive analyzing power 
of the quark distribution will be simply indicated as 
``Sivers asymmetry'', or simply ``asymmetry''. 

\subsection{Sivers-like asymmetry in terms of two-component 
quark states} 

On the theoretical side, $q(x,\vec k_T)$ may be 
defined as the $\gamma_+-$projection of the two-point correlation 
function $C(x,\vec k_T)$ 

\begin{equation}
q(x,\vec k_T)\ =\ Tr[\gamma_+ C(x,\vec k_T)],
\end{equation}
\begin{equation}
C(x, \vec k_T)\  
=\ \int d\xi d^2 b\ e^{- i x \xi + i \vec k_T\cdot \vec b} 
<P|\Psi(0) \bar{\Psi}(0_+,\xi, \vec b)|P>\nonumber
\label{eq:correlator1}
\end{equation}
\begin{equation} 
\propto\ 
\sum_n |\psi_n(x,\vec k_T)|^2. 
\label{eq:sivers2}
\end{equation} 

In this work, the strategy will not be a direct calculation of 
$C(x,\vec k_T)$ from the definition eq.\ref{eq:correlator1}, but a 
calculation 
of a restricted number of amplitudes $\psi_n(x,k_T)$ in 
eq.\ref{eq:sivers2}. 
These are later squared and summed to obtain the correlator. 
This 
choice is associated with 
the need of introducing 
ISI that have continuous and nonperturbative character.  

The functions $\psi_n$ $=$ $<n|\hat q(x,\vec k_T)|hadron>$ 
must be read as quark distribution amplitudes, i.e. 
amplitudes for removing 
a quark with quantum numbers $x, \vec k_T$ 
from the initial hadron leaving the spectator in the state 
$|n>$.\footnote{The 
definition of $q(x,\vec k_T)$ in terms of 
the correlation amplitude is present in several works since 
\cite{CollinsSoperSterman} at least. For more 
details on its translation in terms of $\sum_n |\psi_n|^2$ 
see e.g.\cite{Bacchetta_phd} or \cite{Mulders99}. 
} 
In any single-particle model for a set of 
bound quark wavefunctions, the above 
$\psi_n(x,\xi)$ coincides with one of 
the single quark bound state wavefunctions in the infinite momentum 
frame. The joint action of the 
$\gamma_+-$projection, and of the light-cone limit condition 
$z_+$ $=$ 0 (equivalent to integration of the quark wavefunction 
over $k_-$) selects the 
correct subspace in the u.r. limit. 

If the states $\psi_n$ of eq.\ref{eq:sivers2} are written in the 
form of eq.\ref{eq:states1}, exploiting eq.\ref{eq:trace1} I get 

\begin{equation}
q(x,\vec k_T)\ =\ Tr[\gamma_+ C(x,\vec k_T)]
\ \propto
\nonumber
\end{equation}
\begin{equation}
\propto\ \sum_n Tr( \bar{\psi_n}\psi_n \gamma_+)
=\ \sum_n \bigg( |\psi_+(x,\vec k_T)|^2\ +\ |\psi_-(x,\vec k_T)|^2 \bigg)_n. 
\label{eq:sivers3}
\end{equation} 

\noindent
where $\psi_\pm(x,\vec k_T)$ are complex functions (and not 4-spinors 
anymore) representing the projections of the quark wavefunction on 
transverse spin states. Inserting eq.\ref{eq:sivers3} in 
eq.\ref{eq:ap1} the asymmetry may be written 
as 
\begin{equation}
A(x,k_x)\ \equiv\ 
{{\sum_n \bigg( |\psi_+(x,\vec k_T)|^2\ -\ |\psi_-(x,\vec k_T)|^2 \bigg)_n} 
\over  
{\sum_n \bigg( |\psi_+(x,\vec k_T)|^2\ +\ |\psi_-(x,\vec k_T)|^2 \bigg)_n} 
}. 
\label{eq:ap2}
\end{equation} 
\noindent

\subsection{The undistorted quark state}

In this work one state $n$ only is considered, so that 
the spinor $\vec \psi$ $\equiv$ $(\psi_+,\psi_-)$ represents 
the splitting of the (polarized) hadron into a quark with spin 
projection $\pm \hat y$ and spectator in this unique 
state $|n>$. 

To later insert initial state interactions, we need 
to express the quark state in space-time 
representation: 

\begin{eqnarray} 
\vec \psi(x,\vec k_T)\ 
\equiv\ 
\int d\xi d^2b 
e^{-ix\xi} e^{i\vec k_T \cdot \vec b} 
\vec \psi(\xi,\vec b), 
\hspace{0.5truecm}
\vec \psi(\xi,\vec b)\ \equiv\ 
\left(
\begin{array}{cc}
\psi_+(\xi,\vec b) \\
\psi_-(\xi,\vec b) 
\end{array}
\right)
\label{eq:fourier1} 
\end{eqnarray} 

I assume that the state $\vec \psi(\xi,\vec b)$ 
is defined by a given value of the 
the quark total angular momentum $\vec J$ in the hadron rest frame, and 
that in this state a nonzero correlation 
$<\vec S_{hadron} \cdot \vec J_{quark}>$ $=$ $+1/4$ is present. 
In other 
words, the quark $\vec J$ coincides with the 
parent hadron spin. 

A nonzero correlation between the hadron spin and the 
quark total angular momentum is necessary, since a spin-related 
effect is impossible 
if a quark transports no information on the parent hadron spin. 
In absence of ISI this correlation would produce a nonzero transversity, 
but not a 
single-spin asymmetry of naive Time-odd origin because of global 
invariance rules. 

In absence of ISI, we may assume that we 
are able to calculate the Fourier transform eq.(\ref{eq:fourier1}) 
and write it directly in impact parameter representation 
as ($PW$ means ``plane wave'', i.e. ISI-undistorted state)

\begin{equation}
\vec \psi(\xi,\vec b)_{PW}\ \equiv\ \phi(\xi)\phi'(\vec b)
\cdot|J_y\ =\ +1/2> 
\label{eq:groundstate0}
\end{equation}

\noindent 
A state with given $J_y$ may be realized via S and P waves, 
corresponding to the scalar and axial vector 
spectators of \cite{JakobMuldersRodriguez97}. 

In this work I have avoided, as much as possible, the introduction of  
parameters that cannot be constrained by data. 
Since elastic proton-proton data 
may be reasonably fitted using 
a pure S-wave state (see section IV), I have limited myself to 
the state: 

\begin{eqnarray}
\vec \psi(\xi,\vec b)_{PW}\ \equiv\ \phi(\xi)\phi'(b)
\left(
\begin{array}{cc}
1 \\
0 
\end{array}
\right)
\label{eq:groundstate1s}
\end{eqnarray}

\noindent 
where $\phi(\xi)$ and $\phi'(b)$ 
have Gaussian shapes. 

The width $\Delta \xi$ of $\phi(\xi)$ has been chosen comparing 
the $k_T$ $=$ 0  
distribution with the shape of the collinear function $u(x)$ 
as given by MRST\cite{MRST} at the scale $Q^2$ $=$ 16 GeV$^2$. 

After an initial fit performed in absence of ISI, the 
fit has been re-tuned again after ISI had been included. 
The final fit is shown in fig.6, and has been performed with 
gaussian width parameter $\Delta \xi$ $=$ 4.5, 
plus all the later discussed parameters. 

Although the full procedure is recursive, and so no parameter is 
fully independent from the other ones, I may say that 
the fit on the MRST collinear distribution is decisive to 
establish the width of $\phi(\xi)$, within small corrections. 
As discussed in 
section IV, longitudinal parameters cannot be extracted from elastic 
data (this is evident from eq.\ref{eq:glauber1c}).  
On the contrary, the width of 
$\phi'(b)$ and all the rescattering parameters are 
constrained by elastic data (see section IV). 

\section{Insertion of initial state interactions (ISI)} 

\subsection{Basic assumptions of this work}

\noindent
Assumption 1) 

a set of independent quark-quark 
interactions
is responsible for hadron-hadron scattering 
at $E$ $=$ 10-60 GeV, transferred momenta 
$Q_T$ $=$ 1-3 GeV/c. 

\noindent
Assumption 2) 

the same interactions produce initial state 
distortions of the 
wavefunction of a projectile quark passing through or near a 
target hadron in  
hadron-hadron inclusive processes, at the same energy. 

Here 
the word ``quark'' may also  
mean ``antiquark''. 
Individual quark-quark interactions are imagined as 
exchanges of  
multigluon sets, that is not trivial to reproduce via 
resummed perturbative 
calculations. 
Assumption (1) receives support from e.g. the data 
of \cite{Henkes92} for small$-Q_T$ scattering, and should 
be reasonable for $Q_T$ $\approx$ 2$-$3 GeV/c. 
Assumption (2) becomes reasonable if interpreted in non-exhaustive  
sense: interactions deduced from 
elastic proton-proton scattering data 
constitute one of the relevant contributions to ISI in inclusive 
processes. 

\subsection{Basic equations: Scattering between 
a quark and a composite hadron} 

To connect exclusive and inclusive processes, 
here an ultrarelativistic generalization of the Glauber 
formalism is 
used.\footnote{This 
set of techniques begins with \cite{Glauber59}. For a 
discussion of the technique, and a 
review of applications to both hadronic and nuclear 
processes, see \cite{Anisovich}. For a detailed 
example of application to the calculation of wave distortions 
in a non-elastic process, i.e. the case that is closest to 
what is done here, see \cite{Nikolaev95}.} 

In Appendix B below, 
a comparison of the steps leading to the nonrelativistic 
and ultrarelativistic forms of the Glauber-distorted quark 
wavefunction  
is presented, together with a discussion of 
the most relevant high energy corrections\cite{Gribov69}. 

\begin{figure}[ht]
\centering
\includegraphics[width=9cm]{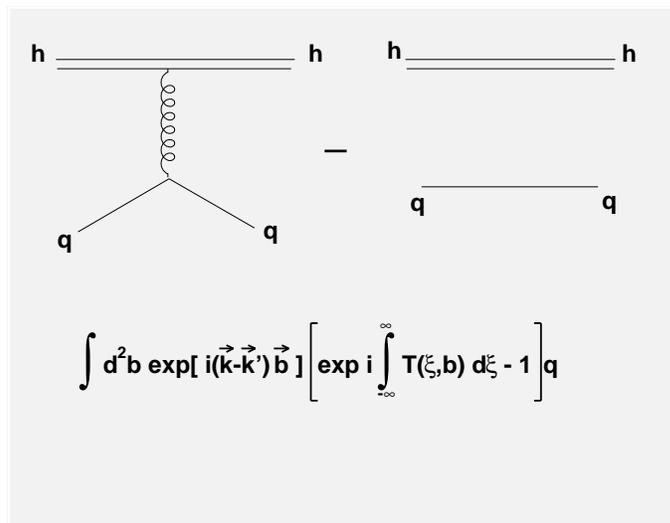}
\caption{quark-hadron scattering amplitude in Glauber u.r. 
approximation. The gluon-shaped t-channel exchange 
reproduces $all$ the terms contributing to the scattering, including 
the no-scattering term that is subtracted on the right side. The 
amplitude for this graph in Glauber approximation (scalar form) is 
proportional to the one in the figure (see 
text and eq.\ref{eq:glauber1}). 
\label{fd_qh_elastic}}
\end{figure}

\begin{figure}[ht]
\centering
\includegraphics[width=9cm]{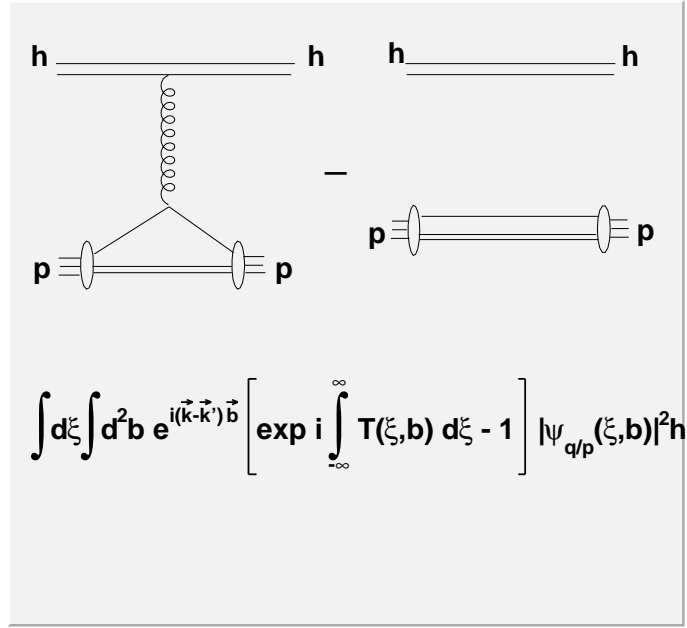}
\caption{proton-hadron scattering amplitude in the same approximation as for 
the previous fig.1. The main difference, evident in the 
(scalar) expression for the amplitude reported in the figure, is 
the average over the quark bound state in the projectile hadron. 
For details, see text and eq.\ref{eq:glauber1c}. 
\label{fd_ph_elastic}}
\end{figure}

\begin{figure}[ht]
\centering
\includegraphics[width=9cm]{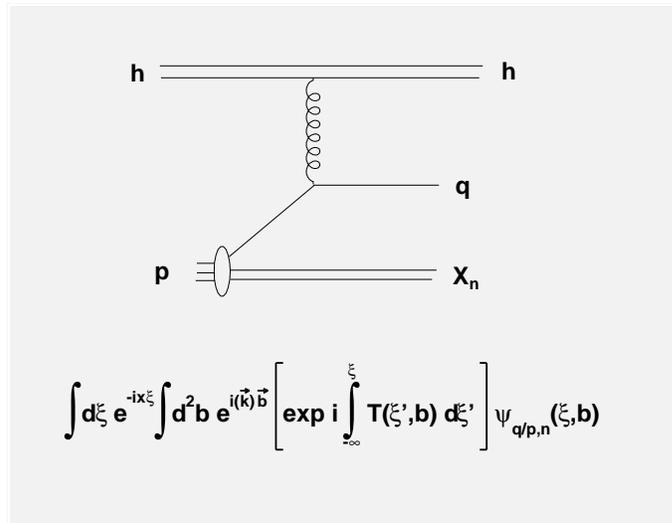}
\caption{The ISI-modified quark distribution amplitude. 
The corresponding (scalar) amplitude is reported in the figure. 
For more precise details, see text and eq.\ref{eq:glauber1d}.
Notice that in this case the ``$-1''$ subtraction is not needed, 
since the no-rescattering term gives a leading contribution 
to the amplitude. 
\label{fd_qd}}
\end{figure}

\begin{figure}[ht]
\centering
\includegraphics[width=9cm]{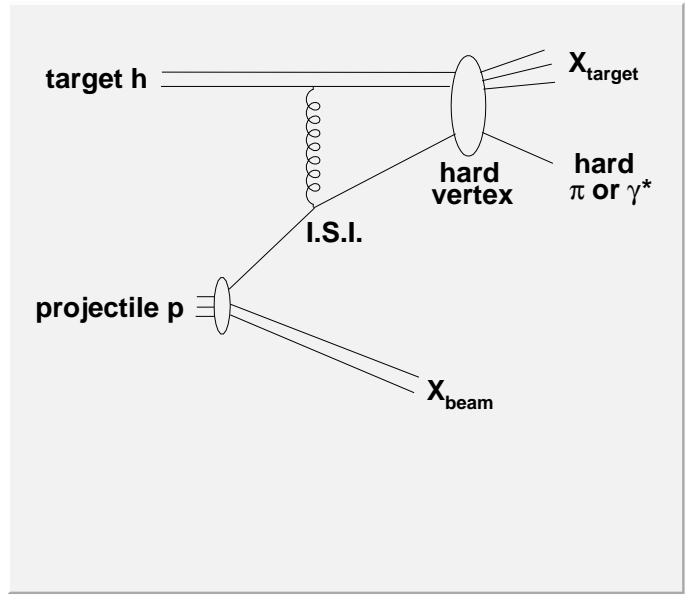}
\caption{The full diagram for inclusive production. 
The blob near the right-upper corner is the true hard interaction 
characterizing the process, also in absence of rescattering. 
The gluon-shaped exchange represents all ISI terms, including 
the no-rescattering one. 
\label{fd_full}}
\end{figure} 

\begin{figure}[ht]
\centering
\includegraphics[width=9cm]{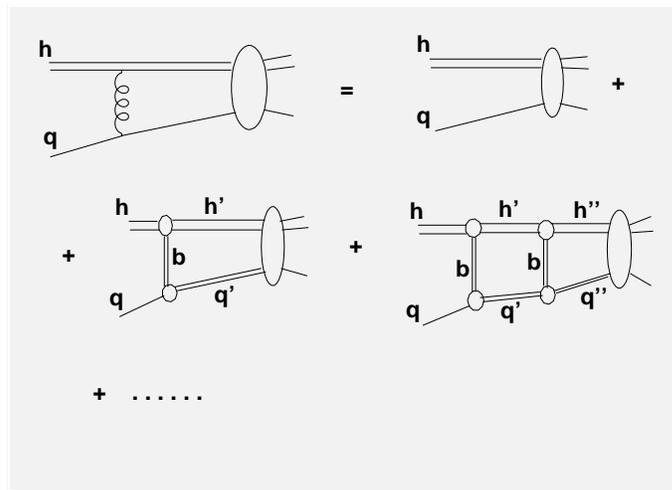}
\caption{The structure of the ISI, attached to the hard scattering blob. 
T-channel double-line exchanges represent non-reducible sets of 
gluon, multigluon, meson, pomeron, etc exchanges. S-channel double 
lines represent intermediate states where the projectile quark 
(or the target hadron) is converted into a more complex intermediate 
state, like $q+g$ or $q+\bar{q}q$. In absence of these intermediate 
excited states the rescattering operator would be hermitean. The 
excited states in the last step (i.e. attached to the hard scattering 
blob) cause a ``direct'' imaginary 
part of the mean field,  
the other ones 
cause an ``optical'' imaginary part (see text, section III.3 
and Appendix B). 
\label{fd_isi}}
\end{figure}

The connection between the amplitudes for describing 
quark-hadron interaction, hadron-hadron interactions, and 
rescattering distortions of the quark distribution amplitude, 
is illustrated in figs.1, 2 and 3. 
The discussion of this section refers to these 
figures. The (slightly) simplified formulas 
shown the figures refer to the scalar 
interaction case, while 
a 2x2 matrix interaction operator is considered in the 
following text and equations. 

As a consequence of assumption (1), 
the elastic scattering amplitude between a free quark 
and a bound quark in a hadron (fig.1) can be written as 

\begin{equation}
T_{el,quark}(k_o,q)\ =\ 
f(k_o) \int d^2 b e^{i\vec q \cdot \vec b} \vec \chi_f^+\Bigg\{
exp\bigg[i \hat P\int_{-\infty}^\infty \hat T(\vec b,\xi)\bigg]\ -\ 1
\Bigg\}\vec \chi_o, 
\label{eq:glauber1}
\end{equation}
\begin{equation}
\hat T(\vec b,\xi)\ \equiv\ \rho(\vec b,\xi) \hat T'(k_o).
\label{eq:glauber1bis}
\end{equation}

\noindent
These relations describe an average of the quark-quark 
scattering amplitude over the target structure. 
In eqs. \ref{eq:glauber1} and \ref{eq:glauber1bis}: 

-) The coefficient 
$f(k_o)$ is a convention-dependent kinematic factor. 

-) $\vec k_o \hat z \mp \vec q$ is the initial/final momentum of the 
free quark, 
$\vec q$ is the transferred momentum, $\xi$ is 
a longitudinal or a light-cone coordinate, $\hat P$ 
indicates path-ordered integration along a constant$-\vec b$ 
light-cone path. 

-) The 2-spinors $\vec \chi_o$, $\vec \chi_f$ 
assign the initial/final spin of 
the projectile quark. These asymptotic spinors are defined $not$ to 
contain the spacetime dependence of the quark wavefunctions. 
In other words, the initial/final wavefunctions for the 
(free) projectile quark 
have the form 
$\vec \chi_{i,f}e^{i(k_o\xi \pm {{\vec q} \over 2} \cdot \vec b)}$. 
(these will have to be generalized later, when we bound the 
projectile 
quark to a projectile hadron).  

-) $\hat T(\vec b,\xi)$ is a 2x2 matrix in the normal spin space.  
$\hat T'(k_o)$ is the 2x2 matrix of $forward$ 
scattering amplitudes between a projectile quark with momentum 
$k_o$ and a target quark at rest. $\hat T'$ depends on $k_o$ only. 
The composition of the target proton in terms of individual 
quarks is contained in 
the single-particle density function 
$\rho(\vec b,\xi)$. 

The target density function $\rho(\vec b,\xi)$ derives from 
an averaging procedure\cite{Glauber59} over all the spectator 
degrees of freedom of the 
target: 
\begin{equation}
exp\bigg[i\int_{-\infty}^\infty d1\ \hat T(1)\bigg]\ 
\sim\ \nonumber
\end{equation}
\begin{equation}
\sim\ \int d2 d3 ....
|\psi_{target}(2,3,...)|^2
exp\bigg[i\int_{-\infty}^\infty d1\ \hat U(1-2,1-3,...)\bigg]\ 
\label{eq:glauber_average}
\end{equation}
Since here the structure of the target only enters through 
the averaged operator $\hat T(1)$, in the following I 
work directly with this operator. For this reason it is 
proper to speak of a ``mean field'' treatment. 

Eq.\ref{eq:glauber1} is pictorially represented by fig.1, 
where 
the ``gluon-shaped'' boson exchange must not be 
meant as a single gluon exchange: it represents the $full$ set of 
possible interactions between a single quark and a hadron, 
approximated by the exponential factor in eq.\ref{eq:glauber1}. 

In 
particular, the exponential factor (represented by the first 
diagram in fig.1) 
also includes the no-interaction term, as 
evident from the fact that it becomes 
unity for $\hat T$ $=$ 0. For this reason, it is necessary 
to subtract the diagonal contribution ``$-1$''. 

\subsection{Basic equations: distortion of the wavefunction 
of a free projectile quark.}

The key point is the possibility to rewrite the first term 
of the right-hand side of 
eq.\ref{eq:glauber1} (i.e. the term that does not contain 
the ``$-1$'' subtraction)
in the form 

\begin{eqnarray}
f(k_o) <\psi_f^{out}(\xi,\vec b) | \psi_o^{in}(\xi,\vec b)>
\equiv\ \nonumber\\ \equiv 
f(k_o) \int d\xi \int d^2 b \ \cdot \ 
e^{-i(k_o\xi - {{\vec q} \over 2} \cdot \vec b)} \vec \chi_f^+ 
exp\bigg[i \hat P\int_{\xi}^\infty \hat T(\vec b,\xi) d\xi\bigg]
\nonumber\\
e^{i(k_o\xi - {{\vec q} \over 2} \cdot \vec b)} 
exp\bigg[i \hat P\int_{-\infty}^\xi \hat T(\vec b,\xi) d\xi\bigg] 
\vec \chi_o
\label{eq:glauber1b}
\end{eqnarray}
so that we may define the concept of ``distorted quark 
wavefunction'', with the $\xi-$dependent distortion factor  
\begin{equation}
F(\vec b, \xi)\ =\ 
exp\bigg[i P\int_{-\infty}^\xi \hat T(\vec b,\xi) d\xi\bigg]. 
\label{eq:glauber2}
\end{equation}

The connection between eqs.\ref{eq:glauber1} and \ref{eq:glauber2} 
is a standard step in the 
Glauber approximation. Within a nonrelativistic 
formalism it was first used in \cite{Glauber59}, 
where also the spin-orbit case was discussed. 

\subsection{Basic equations: hadron-hadron scattering}

Now two relevant generalizations are possible. First, we may substitute 
the wavefunction of a free projectile quark with the wavefunction of a 
quark that is bound to a projectile proton with momentum 
$p \hat z \pm \vec q / 2$ (fig.2). 
In this case the scattering 
amplitude is (apart for an overall kinematic factor) 

\begin{eqnarray}
T_{EL}(p,q) =\ \int d\xi \int d^2 b 
\nonumber \\
\vec \psi^+(\xi,\vec b)
e^{i\vec q \cdot \vec b} \Bigg\{
exp\bigg[i \hat P\int_{-\infty}^\infty \hat T(\xi,\vec b) d\xi\bigg]\ -\ 1
\Bigg\} \vec \psi(\xi,\vec b) 
\label{eq:glauber1c}
\end{eqnarray}
and it describes the scattering between a projectile proton and a 
target hadron, for $q$ $<<$ $p$. 
In the following, this amplitude is used to fit a reasonable 
form for the $\hat T$ operator starting from data on angular distribution 
and normal analyzing power in 
elastic proton-proton scattering at beam energies 20-50 GeV. 

The role of the ``$-1$'' subtraction must be stressed. It means that 
the leading scattering term is part of the interaction 
eikonal operator of eq.\ref{eq:glauber1c}. 
In the limit of no interaction the above 
amplitude is zero. 
If the ``$-1$'' factor is removed, in the limit of no interaction 
the above equation gives the nonperturbative expression 
of the projectile form factor 
$<\psi(p+q/2)|\psi(p-q/2)>$. In this case the eikonal operator may 
only contain second order corrections to the form factor, and the 
``hard'' features of the scattering event, if present, must 
be inserted into the large-momentum tail of the bound state 
wavefunction. In this 
work we adopt the ``$-1$'' subtraction, suitable to describe 
semihard elastic scattering with $q$ $<<$ $p$.  This means to assume 
that the bound state itself does not contain hard momentum tails. 

\subsection{Basic equations: DWBA and the modifications of the quark 
distribution functions.}

The other useful generalization is 
the application of the distortion factor to the calculation of the 
quark distribution function, illustrated in fig.3. It 
exploits the Glauber approximation within 
the Distorted Wave Born Approximation (DWBA) scheme. In DWBA a 
matrix element of the form $<f|(A+B)|i>$, with $A$ and $B$ 
interaction operators, is written as $<f'|B|i'>$, where  
$|i'>$ and $|f'>$ are solutions of the problem where 
the interaction $B$ is excluded while $A$ is considered. 
This allows for a compared study of different processes  
whenever one may assume that the underlying distorting factors $A$ 
have the same 
origin. The procedure is of course justified if 
$B$ is harder than 
$A$.\footnote{For 
a detailed example see \cite{BREikonal}. Here,  
high-energy 
predictions for electron and proton scattering on a nucleus $A$, i.e.  
for the processes $A(p,p)A$, $A(e,e'p)A-1$, and $A(p,2p)A$, are 
related within DWBA, and the Glauber-style distorted wavefunction is 
compared with distortions calculated by other 
methods.} 

The ISI-distorted quark distribution amplitude is 

\begin{equation}
\vec \psi_n(x,\vec k_T) =\ \int d\xi \int d^2 b 
e^{-ix\xi + i\vec k_T \vec b} 
exp\bigg[i \hat P\int_{-\infty}^\xi \hat T(\xi,\vec b) d\xi\bigg]
\vec \psi_{n,PW}(\xi,\vec b) 
\label{eq:glauber1d}
\end{equation}
where $\vec \psi_{n,PW}(\xi,\vec b)$ is the amplitude for 
extracting a light-cone quark in position 
$(z_-,z_+,\vec z_T)$ $=$ 
$(P_+\xi,0,\vec b)$ 
while leaving the spectator in 
the state $|n>$, in absence of ISI. 
In this work $\vec \psi_{n,PW}(\xi,\vec b)$ 
coincides 
with the 
quark bound state given by eq.\ref{eq:groundstate1s}. 
For a generic nonzero $\hat T(\xi,\vec b)$,  
eq.\ref{eq:glauber1d} defines the quark distribution amplitude that 
enters eqs. \ref{eq:sivers3} and \ref{eq:ap2} in presence of ISI. 

It must be noted that the ``$-1$'' factor is missing in 
eq.\ref{eq:glauber1d}, since 
the undistorted term is a leading contribution in this 
case. In other words, for $\hat T(\xi,\vec b)$ $=$ 0 
the right hand side of eq.\ref{eq:glauber1d} is anyway nonzero 
and coincides with the ISI-not-affected quark distribution 
amplitude $\vec \psi_{n,DW}(\xi,\vec b)$. 

As in figures 1 and 2, 
the gluon-shaped factor of fig.3 represents the eikonalized 
full set of rescattering interactions. Since we are speaking 
of ``rescattering'', these  do $not$ include the 
true hard scattering event defining the process class 
(Drell-Yan, meson production, etc). To stress this point, in fig.4 
I show the amplitude for the full process interesting here.  
The ``strictly hard'' event is contained in the upper-right blob 
from which a hard line representing a jet, a meson or   
a massive photon emerges. The gluon-shaped ISI 
include everything softer than the final hard event. 

\subsection{Complex mean field}

With the aim of arriving to a nonzero Sivers asymmetry, it 
is important that the rescattering operator contains different 
terms able to introduce phase shifts between competing 
amplitudes for the same process: $\hat T$ $=$ $\sum \hat T_i$. 
A special role in 
this respect is played by anti-hermitean 
interaction terms. I will simply speak of ``real'' and ``imaginary'' 
terms to mean hermitean and anti-hermitean 2x2 interaction 
matrices $\hat T_i$. 

In fig.5 the gluon-shaped ISI of fig.4 are expanded in terms 
of non-reducible t-channel exchanges. 
Intermediate s-channel states that in fig.5 are 
named $q'$, $q''$, etc, are in 
general more complex than quark states 
(e.g. a quark may split into quark+gluon, and the gluon 
may be absorbed in the next vertex). The presence of 
imaginary terms is related with the additional 
cuts that may be applied to these intermediate states. 

The imaginary terms   
may have two 
origins\cite{Feshbach,Glauber59,Gribov69}: direct and optical.   

The direct one is the case where 
the underlying quark-quark scattering amplitude $\hat T'$ is 
complex, and the process 
is single-scattering dominated. Since the described mean field  
is an average of $\hat T'$ over the target matter distribution, 
a complex $\hat T'$ leads to a complex mean field. This mechanism 
is surely present in the problem under consideration, and corresponds 
to the single rescattering graph in fig.5 when $q'$ 
$\neq$ $q$ or $h'$ $\neq$ $h$. 

The optical case requires a two-step transition, and is 
the case where  
the initial state is regenerated after 
passing through an intermediate inelastic channel. Then, the 
existence of a cut in the intermediate state 
produces the imaginary part. This corresponds to taking the 
last contributing diagram in fig.5 with $q''$ $=$ $q$, 
$h''$ $=$ $h$, but $q'$ or $h'$ different from $q$ or $h$. 
The effective imaginary part is then present in 
$\rho(b,\xi)$. This optical effect may 
be defined of Gribov's kind\cite{Gribov69} (see Appendix B). 
The well-known 
optical effect of Feshbach's kind\cite{Feshbach} 
implies the formation of bound $q+h$ states but is 
suppressed in the short wavelength regime considered here.  

A complex  
interaction term implies flux nonconservation. 
Obviously one expects absorption of flux from the initial 
state. To conserve unitarity, this lost flux will be 
redistributed among all the other accessible states. 
This implies opposite-phase imaginary terms dominating at small 
and large angles in the elastic channel. At small angles, 
we have some elastic scattering due to diffraction, but 
the net effect on the incoming flux is absorption, 
meaning a negative imaginary part for the interaction 
term. At large angles, second order transitions 
``elastic $\rightarrow$ inelastic $\rightarrow$ elastic''
produce elastic flux. In a single channel approach, this is 
reproduced by a complex interaction term with 
positive imaginary part. The signature of this is the 
minimum in the data of fig.7 at $Q_T$ $=$ 1.2 GeV/c. 


\subsection{Selected interaction terms} 

The determination of $\hat T$ $\equiv$ 
$\rho(\xi,b)\hat T'(k_o)$ 
entering eqs. \ref{eq:glauber1},\ref{eq:glauber1bis} 
is not unique, since different $\hat T(b,\xi)$ 
may lead to the same $\hat T_\infty(b)$ $\equiv$ 
$\int_{-\infty}^{\infty} \hat T(b,\xi) d\xi$. In addition, 
this operator derives from the averaging procedure 
eq.\ref{eq:glauber_average} that is under control for simple 
few-body systems only (see \cite{BJNZ} for a borderline 
example). 
So, in cases like the one interesting here 
one is obliged to 
guess and fit directly some model form for $\hat T(\xi,b)$. 

I assume that the distorting factor 
$\hat F$ of 
eq.(\ref{eq:glauber2}) 
does not present a fast dependence on $x$, so in fourier transforms 
in may be considered as $x-$independent. This assumption breaks down 
at small $x$, but that region is of no 
interest here since it clearly involves a different physics 
under several points of view. 
In numerical calculations, the path-ordered exponential 
operator is approximated by a quasi-continuous product: 

\begin{equation}
exp\Bigg( i \hat P \int_{-\infty}^\xi \hat T(\xi',\vec b) d\xi' \Bigg)
\ \approx\ 
\prod (1 + i \hat T d\xi)
\end{equation}
where the product starts from a negative and large enough $\xi'$ 
value where interactions may be neglected, and stops at $\xi$. 
The $\hat T$ matrix is a sum of the kind 

\begin{equation}
\hat T\ =\ \sum \hat T_{scalar}\ +\ \sum\ \hat T_{spin-orbit}. 
\end{equation}

where two scalar ant two spin-orbit terms are included, and 
each term has the form: 

\begin{eqnarray}
\hat T_{scalar,n}\ \equiv\ 
\delta_n \left(
\begin{array}{cc}
1  &  0 \\
0 &  1 
\end{array}
\right)
\rho(\xi)\rho_n'(b).
\label{eq:delta_n}
\end{eqnarray}

\begin{eqnarray}
\hat T_{spin-orbit,n}\ \equiv\ \alpha_n \left(
\begin{array}{cc}
b_x &  i b_y \\
-i b_y &  - b_x 
\end{array}
\right)
\rho(\xi)\rho_n'(b).
\label{eq:alpha_n}
\end{eqnarray}

\noindent
All the density functions have been chosen with gaussian form, 
normalized to $\rho_i(0)$ $=$ 1. 
The longitudinal density $\rho(\xi)$
is the same for 
all terms and is equal to the squared quark  distribution amplitude  
$|\phi(\xi)|^2$ introduced in eq.\ref{eq:groundstate1s}. 
In the following 3 transverse density gaussian functions 
$\rho_n'(b)$ will 
be needed: a soft, a semi-hard and a ultra-hard one. In all, 
I will introduce a soft and a semi-hard scalar term, and 
a soft and a ultra-hard spin-orbit term. 
The $\delta_n$ coefficients are complex, while the $\alpha_n$ have been 
chosen as pure imaginary. See next section for details on the fitting 
procedure and for their values. 

To see that the $O(\alpha_n)$ terms 
are spin-orbit terms, I remark that the above matrices 
act on a basis of eigenstates of $\hat S_y$. This means that 
$\hat S_y$ $=$ $\sigma_3$, $\hat S_z$ $=$ $\sigma_1$, $\hat S_x$ 
$=$ $\sigma_2$ (apart for a factor 1/2). 
So we may rewrite: 

\begin{equation}
\hat T_{spin-orbit}\ d\xi\ =\ \rho\rho' \alpha \Big(b_x \hat S_y\ 
-\ b_y \hat S_x \Big) P_+ dz_-  
\label{eq:T1}
\end{equation}

Since $k_z$ $\propto$ $xP_+$, we identify the triple product 
between $\vec b$, $\vec k$, and $\vec S$ in the previous equation, and 
this is equal to $\vec L \cdot \vec S$. 

\begin{figure}[ht]
\centering
\includegraphics[width=9cm]{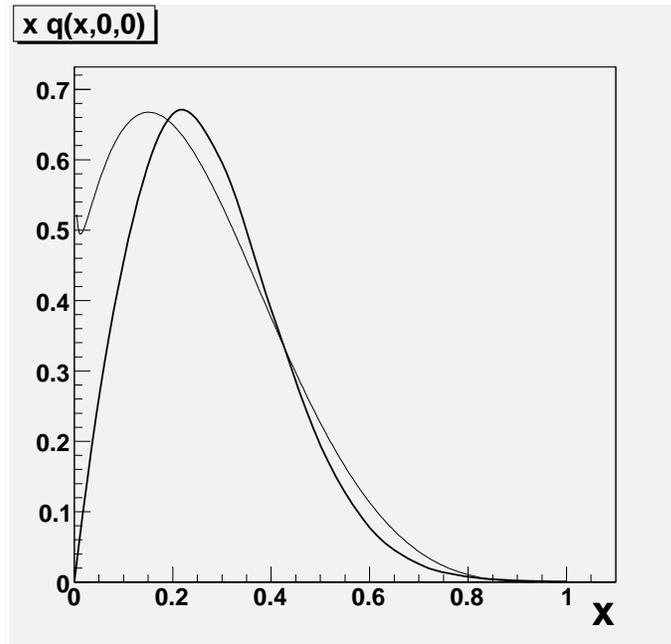}
\caption{Unpolarized quark distribution: MRST\cite{MRST} collinear 
distribution $x u(x)$
at $Q^2$ $=$ 16 GeV$^2$ and model $x q(x,k_T=0)$. The model curve is 
the one that tends to zero at small $x$. 
\label{res_sidis_qx}}
\end{figure}

\begin{figure}[ht]
\centering
\includegraphics[width=9cm]{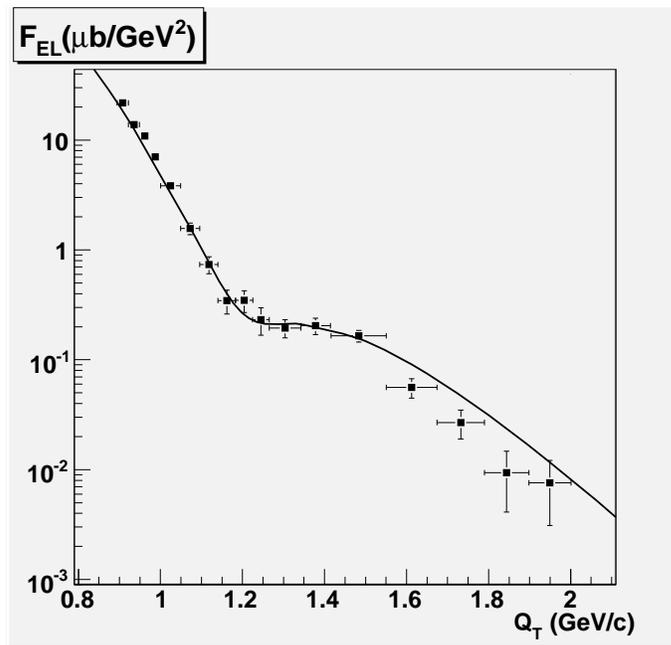}
\caption{Fit of elastic $pp$ scattering data. $F_{EL}$ $\equiv$ 
$d\sigma/dt$. Data come from ref.\cite{Asad85}. 
\label{res_elastic_f}}
\end{figure}

\begin{figure}[ht]
\centering
\includegraphics[width=9cm]{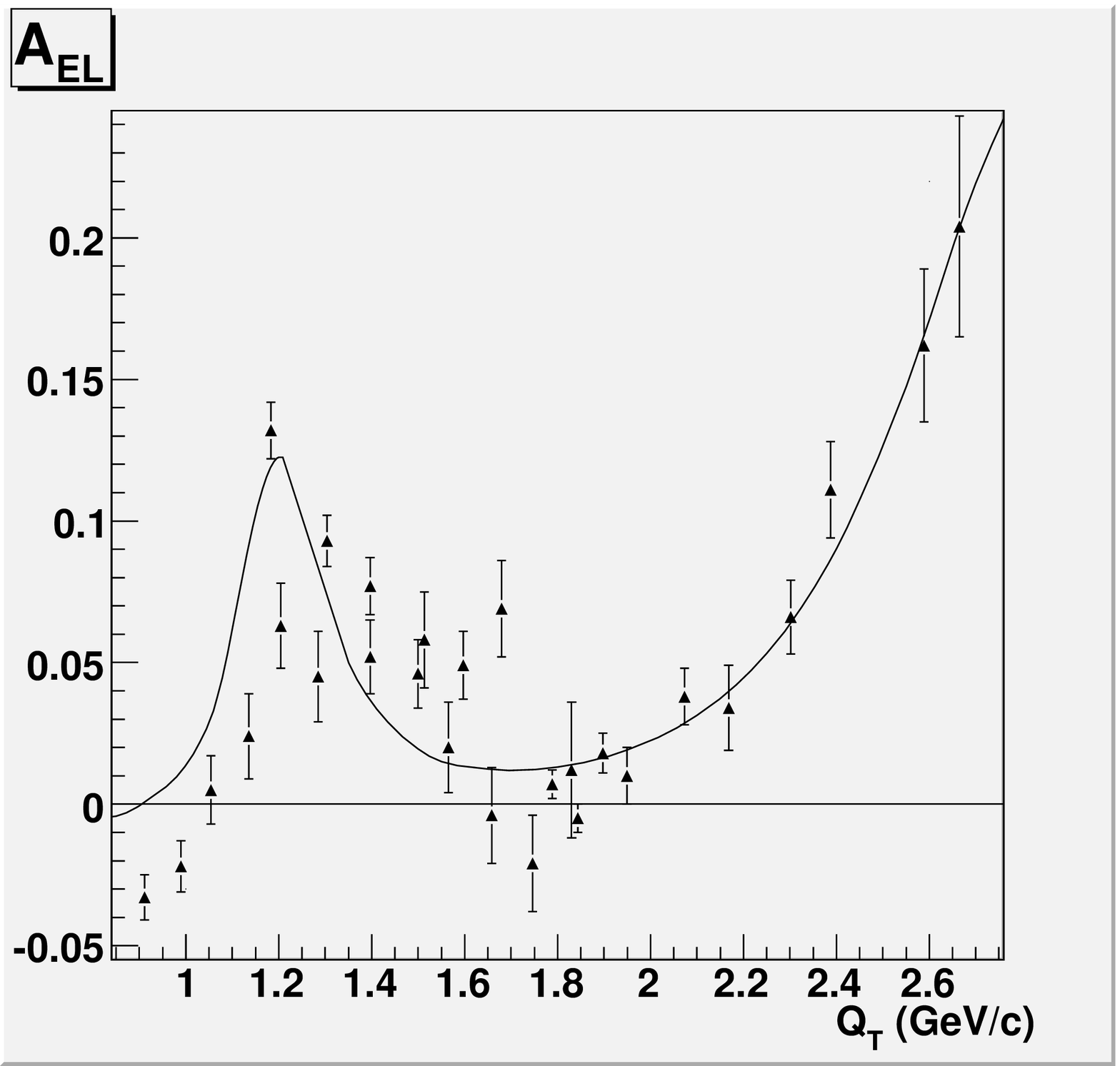}
\caption{Fit of the analyzing power in elastic pp data. 
$A_{EL}$ $\equiv$ $[F_{EL}(k_x)-F_{EL}(-k_x)]/[F_{EL}(k_x)+F_{EL}(-k_x)]$.  
Data come from refs.\cite{Crabb90,Cameron85,Peaslee83,Hansen83,
Antille81}. 
\label{res_elastic_ap}}
\end{figure}

\section{Fit of scalar and spin-orbit interaction terms.}

For extracting the distortion factor eq.\ref{eq:glauber2} from 
eq.\ref{eq:glauber1c} applied to proton-proton elastic scattering, 
I have assumed that the projectile hadron participates 
to both processes (SIDIS or elastic scattering) with the 
same quark ``intrinsic'' distribution amplitude 
eq.\ref{eq:groundstate1s}.

The phenomenology of normal spin observables in  
proton-proton scattering 
at 20-50 GeV, and $Q_T$ $=$ 1$-$3 GeV/c,  
is not trivial. 
Whichever the model, more analogous terms must be summed to 
reproduce it. 
Indeed, data 
show interference between at least two competing 
terms in elastic scattering\cite{Asad85,LLL93}, and three 
competing terms in single-polarization 
measurements\cite{Crabb90,Cameron85,Peaslee83,Hansen83,Antille81}. 

Here I ignore the behavior of the analyzing power in the 
very soft region (where it is nonzero but small), and include 
two scalar complex terms, and two imaginary  
spin-orbit terms.
According to the relations of the previous sections, 
in all the cases any of the interaction 2x2 terms contains an overall 
scalar factor of the form $(\beta_R + i \beta_I) 
exp[-(b/\Delta b)^2]exp[-(\xi/\Delta \xi)^2]$, where $(\beta_R + i \beta_I)$ 
is the strength 
parameter, $\Delta b$ the transverse range and $\delta \xi$ the longitudinal 
range. 

Also, the parameters of the quark bound state in the projectile 
are relevant, since this state is convoluted with the 
target interaction operator in the relevant matrix elements. 
This state too contains a factor of the form 
$exp[-(b/\Delta b)^2]exp[-(\xi/\Delta \xi)^2]$, 
so we may speak of transverse radius and longitudinal range for 
the bound state. 

As evident from eq.\ref{eq:glauber1c} 
the data of figs.7 and 8 cannot give information on any of 
the longitudinal ranges, of the quark bound state or of the interaction 
terms. For the quark bound state this parameter is 
constrained by the comparison with collinear distribution 
functions\cite{MRST} (fig.6). 

I have taken the 
longitudinal range of the quark bound state equal to 4.5, and 
all the longitudinal 
ranges of the interaction terms equal between them and equal 
to $4.5/\sqrt{2}$. This means that for all densities 
we have $\rho_i(\xi)$ $=$ $|\phi(\xi)|^2$. In other words, the 
same longitudinal range is attributed to all the terms of interest 
here. 

This longitudinal 
range has been fixed by the fit in fig.6 
a first time by calculating the unpolarized distribution in absence 
of all rescattering terms. It has been fine-tuned 
a second time after including the leading soft scalar ISI 
term (see below). It has been tuned once again 
after the remaining ISI terms had been included, but with no effect.  
 
Data in fig.7 reproduce $pp$ elastic scattering at beam energy 50 GeV, 
taken from ref.\cite{Asad85}. 
The fit curve in 
this figure shows the left/right-averaged $Q_T-$distribution, 
i.e. for each $Q_T-$value I report the average of the 
two values of the distribution corresponding to $Q_T = \pm |Q_T|$.   
After this average, spin-orbit terms have negligible 
effect on the fit.\footnote{They 
have small but nonzero effect. Indeed, 
the left-right asymmetry is due to the interference between spin-orbit 
and scalar terms. Squared spin-orbit terms produce $Q_T-$even 
contributions, that in the present case are small for $Q_T$ up to  
3 GeV/c.} 

So spin-orbit terms 
are constrained by the data in fig.8 only. This figure reports 
data points from several experiments measuring the proton-proton 
normal spin analyzing power at 20-30 GeV beam 
energy\cite{Crabb90,Cameron85,Peaslee83,Hansen83,Antille81}. 

\subsection{The soft scalar term.}

In fig.7 two regions are evident: a soft and a semihard region, 
corresponding to $Q_T$ below and over 1.2 GeV/c. The region below 
1 GeV/c is dominated by the soft scalar term. 

There is an important constraint on this term. For forward 
scattering, (i) the ratio $\rho$ of the real to the imaginary part  
of the amplitude is measured and assumes values between 
$-5$ and $-30$ \% (decreasing with energy) in the 
region 20-50 GeV\cite{Bellettini65,LLL93}, (ii) we are in 
single scattering regime, and the use of a Born-1 approximation 
is justified.  
This fixes the ratio of the real to the imaginary part  
of the related interaction potential. 
Indeed, 
\begin{equation}
A_{born,1}(Q=0)\ \propto\ \int \hat V(\xi,\vec b) d\xi d^2b
\end{equation}
and 
\begin{equation}
V(\xi,\vec b)\ \propto\ exp\bigg(
\int_{-\infty}^\xi \hat T(\xi',\vec b) d\xi'
\bigg) - 1\ \approx\ 
\int_{-\infty}^\xi \hat T(\xi',\vec b) d\xi',  
\end{equation}
So, for the soft scalar term that dominates forward 
scattering, $\hat T(\xi,b)$ has the same phase as $\rho$. 

I have taken $Re(T)/Im(T)$ $=$ $-0.2$ (averaging a decreasing 
trend in the range 20-50 GeV). In practice, 
this ratio has little influence on the following, 
I have just assumed this value to conform with the known part of the 
phenomenology. 

Since the central data peak is due to peripheral events, I have 
assumed that the transverse range of the interaction is 
much larger than the transverse radius of the bound quark state. 
With this assumption, the strength and slope of the central peak 
fix at once the strength of the scalar 
soft interaction term 
and the transverse radius of the bound state of the projectile 
quark. The transverse radius of the bound state is 1 fm. 
The strength of the scalar soft term is reported below, 
together with the strength of all terms. 
The precise value of the transverse range 
of the soft term has no effect as far as it is much larger 
than the bound state radius. I have taken it equal to 4 fm, but 
with 3 or 7 fm things do not change. 

More in general, for any interaction term the (transverse) integral 
of 
eq.\ref{eq:glauber1c} is cut off by the shorter between the 
bound state transverse and the interaction transverse 
range.
When there is a relevant difference between the magnitude of the 
two, the precise value of the larger is not relevant. 
It must however be remarked that this property does not 
transfer automatically to 
eq.\ref{eq:glauber1d}, because of the absence 
of the ``$-1$'' subtraction in that case. This is clearly a 
source of ambiguity. In the case of the soft term it is anyway 
licit to guess that it has little effect on the 
asymmetry of the inclusive distributions, 
because of its reduced range in $Q_T-$space. 

\subsection{The semihard scalar term.}

The parameters of the semihard scalar interaction term are 
given 
by fitting unpolarized scattering data (fig.7) in the shoulder region at 
$Q_T$ $>$ 1 GeV/c, 
and by the need of obtaining a nonzero 
analyzing power at $Q_T $ $>$ 1 GeV/c (fig.8). 

The real and imaginary parts of the semihard scalar term are 
assumed 
equal in modulus. The sign of the imaginary part corresponds to 
flux production, 
and the sign of the real part to repulsive interaction. 
The transverse range of this term is  
fixed to 0.45 fm 
by the shape of the data of fig.7 for $Q_T$ $>$ 1 GeV/c. 

A qualitative consideration of this semihard scalar term suggest 
that it mimics, within a single channel formalism, the 
effect of double scattering terms with intermediate formation 
of large-mass states (see the related discussion in section III.3  
and Appendix B). 

For fitting purposes, a nonzero imaginary part is 
needed, with opposite sign with respect to the soft term, 
to reproduce the interference pattern at 1.2 GeV/c. 
A nonzero positive real part does the same, because of 
a similar interference with the real part of 
the soft term. Because of the real to imaginary ratio  
of the soft part, with these signs the real part of the 
semihard potential contributes to 20 percent of the 
interference effect in the dip region. 

In presence of a zero imaginary part, 
one would require a much larger 
strength for the real part to reproduce the dip. 
This larger strength 
would not fit the region $Q_T$ $>$ 1.2 GeV/c. 
The same would happen for a negative real part: the imaginary 
part should be increased to compensate. 

The major role of the real part is to 
interfere with the imaginary spin-orbit terms producing a nonzero 
analyzing power. So, the real constraint on the real part arrive 
from the joint fit of data of fig.7 and fig.8. In addition, 
the strengths of the real part of the semihard scalar term and of the 
hard spin-orbit term are not fully independent.  

\subsection{Spin-orbit terms}

Analyzing power data present three characteristic regions, 
only two of which evident in fig.8: (i) a low but nonzero 
(positive) broad peak 
at very small $Q_T$ (not showed in fig.8, but visible e.g. in 
\cite{Crabb90}); (ii) a peak between 1 and 1.8 GeV/c; 
(iii) a large increase over 1.8 GeV. 

Here, I neglect the ultra-soft peak at small $Q_T$. I introduce two 
spin-orbit imaginary terms: a soft term, with the same density 
structure as the scalar soft term, and a hard term, with a 
very small transverse interaction range 0.16 fm. 
The interference between the former and the semihard scalar term 
(real part) produces the nonzero values in the region 
1$-$1.8 GeV/c, while the interference between 
the latter and the semihard scalar term (real part) produces 
the large$-Q_T$ rise. 

The strength parameter for the soft spin-orbit term has 
been tuned so to best reproduce those points that present the 
smallest error bars in the confused data set of 
the region 1$-$1.8 GeV/c. 

The hard spin-orbit term is necessary to reproduce the increase 
of the analyzing power in the $Q_T$ $>$ 1.8 GeV/c region. 
In fig.8, some points at large $Q_T$ with 
large error bars have not been reported. If taken at 
their central value, they would suggest even larger analyzing powers 
than the fitted ones. In addition, my fitting curve decreases 
for $Q_T$ $>$ 3 GeV/c. The decrease is driven by the 
fall of the scalar semihard term interfering with 
the spin-orbit terms. But we do not know what 
happens to data at $Q_T$ $>$ 3 GeV/c. 

The dip at 1.6$-$1.8 GeV/c is due to the fact that neither of 
the two introduced spin-orbit terms is strong there. 
I have $not$ introduced a 
semihard spin-orbit term with range 0.5 fm as for the scalar 
potential, since it would produce 
large analyzing powers at $Q_T$ $=$ 1.6$-$1.8 where we see 
small or even negative analyzing 
power. Some of the small-error points reported in the figure 
are born from measurements that were specifically 
dedicated to the region 1.6$-$1.8 GeV/c. They confirmed with little 
doubt that in this region the analyzing power is close to 
zero. 

The other dip, at 1 GeV/c, is well measured, and is due to the 
cancellation between the real parts of the two scalar 
interactions. Then the imaginary spin-orbit soft term cannot 
interfere with anything. On the left of this dip my fitting 
curve produces a very small, negative, rather flat analyzing 
power that does not correspond to reality 
(see e.g. \cite{Crabb90}).  
Data show an again positive, but smaller 
than 5 \%, analyzing power in the most forward region 
$Q_T$ $<$ 0.5 GeV/c. 
So, an extra term 
would be needed to explain quasi-forward 
data. 
Because of the smallness of the effect, 
I have not cared data at $Q_T$ $<$ 1 GeV/c. 

\subsection{Parameter values}

Bound state longitudinal range: $\Delta \xi$ $=$ 4.5.  

Longitudinal range of all the density widths: 
$\Delta \xi$ $=$ 4.5/$\sqrt{2}$.  

The transverse width is peculiar: 

quark bound state: $\Delta b$ $=$ 1 fm. 

soft scalar and soft spin-orbit interaction: $\Delta b$ $=$ 4 fm. 

semihard scalar interaction: $\Delta b$ $=$ 0.45 fm. 

hard spin-orbit interaction: $\Delta b$ $=$ 0.16 fm. 

The strength parameters are:

soft scalar interaction: $\delta$ $=$ 0.07 (i-0.2). 

semihard scalar interaction: $\delta$ $=$ 0.01 (1-i). 

soft spin-orbit interaction: $\alpha$ $=$ 0.0004 i. 

hard spin-orbit interaction: $\alpha$ $=$ 0.0001 i. 

\section{Estimate of effects in inclusive processes} 

\subsection{Distributions and asymmetries at quark level}

Fourier transforms have been calculated assuming 
$\vec k_T$ coinciding with $(k_x,0)$. 

Fig.9 shows the quark 
distribution for positive and negative $k_x$, for $x$ $=$ 0.3. 
The two curves of fig.9 may also be read as the distributions 
corresponding, in both cases for positive $k_x$, to a parent 
hadron with spin oriented as $\pm \hat y$. Their non-identity implies 
a Sivers-like asymmetry and a nonzero function $q_S(x,k_T)$ 
according with eq.\ref{eq:sivers}: 
$q(x,\vec k_T)$  
$\equiv$ $q_U(x,k_T)$ $+$ $(k_x/M) q_S(x,k_T)$, 
for an unpolarized quark in a hadron with polarization along 
$+\hat y$. For opposite hadron polarization $(k_x/M)q_S$ is 
substituted by $-(k_x/M)q_S$ (see eq.\ref{eq:sivers}). 
The $q_S/q_U$ ratio extracted from the difference between 
the two curves in fig.9 
is shown in fig.10 as a function of $k_T$.

\begin{figure}[ht]
\centering
\includegraphics[width=9cm]{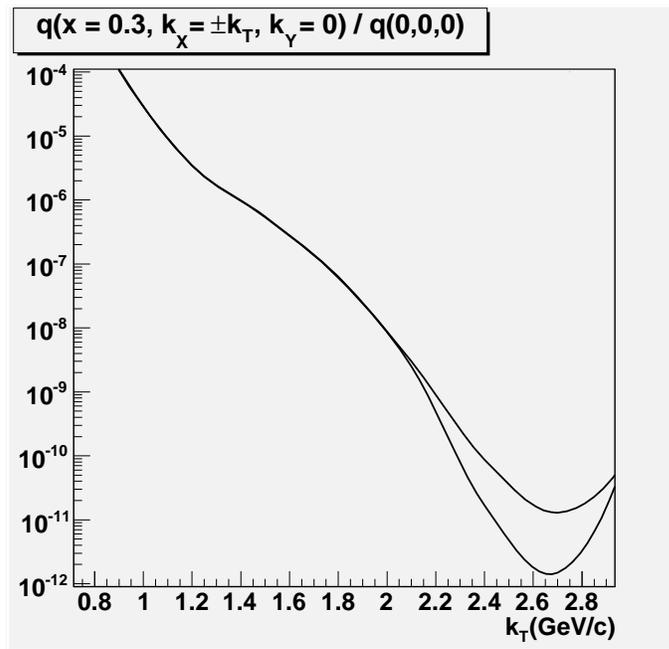}
\caption{$k_x-$dependence for $x$ $=$ 0.3 and $k_y$ $=$ 0 
of the 
quark distribution $q(x,k_x,0)$ (see eq.\ref{eq:sivers}
in the text) for positive and negative $k_x$. 
These two curves refer to a parent proton polarized 
along $+\hat y$ and to an unpolarized quark. However, 
they also represent the positive $k_x$ distribution in the 
case of parent protons with opposite polarization $\pm \hat y$. 
The asymmetry between the shown curves is a Sivers-like 
asymmetry, from which the ratio between the Sivers and 
the unpolarized distribution may be extracted (next figure). 
\label{res_sidis_qkt}}
\end{figure}

\begin{figure}[ht]
\centering
\includegraphics[width=9cm]{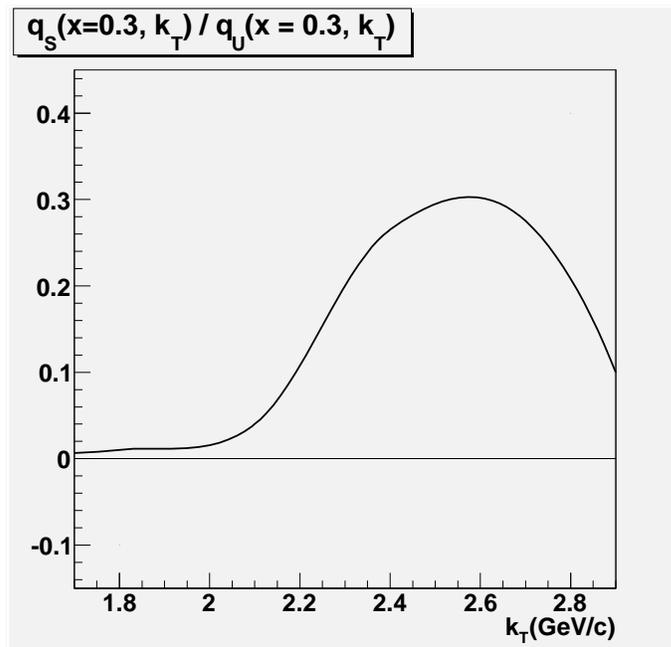}
\caption{Ratio between the Sivers and the unpolarized 
distribution functions $q_S(x,k_T)$ and $q_U(x,k_T)$, 
as extracted from the asymmetry of the two curves reported  
in the previous figure (see eq.\ref{eq:sivers} in the text). 
For $k_T$ $<$ 1.8 GeV/c the curve does not overcome 0.02. 
\label{res_sidis_ap}}
\end{figure}

Observations:

1) Figs 9 and 10 are completely representative of what may be seen at 
other values of $x$ in the valence region, e.g. $x$ $=$ 0.2. 

2) Repeated application of the numerical calculation code shows that 
numerical calculations lose progressively reliability for $k_T$ $>$ 
2.7 GeV/c at $x$ $=$ 0.3. For larger $x$ this loss of reliability 
is more severe, because of the fourier factor $exp(ix\xi)$. For 
this reason the value $x$ $=$ 0.3 has been chosen, as the 
largest $x$ value for which a reasonable $k_x-$range could 
be covered. For the same reason, $k_y$ has been fixed 
to zero. 

Observing fig.9, although a flattening of the 
distribution at $k_T$ over 2 GeV/c is to be expected, the steep 
rise of the 
$k_T$ distribution at $k_T$ $\approx$ 3 GeV/c is not 
reliable. 
Up to 2.7 GeV/c the results of the numerical calculations are 
stable against changes of the integration point numbers and/or 
integration ranges. In the case of the figures referring 
to elastic scattering, stability arrives to 3 GeV/c. 

3) The peak of the asymmetry evident at 2.4 GeV/c in fig.10, 
together with the disappearance of the asymmetry over 3 GeV/c,  
is not 
an effect of the code unreliability, but of the chosen interaction 
shapes. In other words, although numerical predictions are not 
reliable at 3 GeV/c, the fourier transform 
of the involved interaction terms must decrease in this 
region. 

4) If one wants to compare the shape of the curves in figs. 9 and 10 with 
those in figs.7 and 8, one 
must observe that  the transverse momentum scale of any phenomenon 
is slightly reduced in the 
quark distribution case.  This is due to the fact that 
in eq.\ref{eq:glauber1c} we find $|\psi(\xi,b)|^2$, while in 
in eq.\ref{eq:glauber1d} we have $\psi(\xi,b)$. Because of the 
chosen gaussian shape, this means slightly harder effects 
in the elastic case. 

5) Of the interaction terms inherited from elastic scattering, only 
the semihard and hard ones produces relevant effects on the quark 
distribution side. 
Soft scalar and spin-orbit terms are necessary to explain 
elastic data, but their effect on the asymmetry in the case of 
inclusive processes is small. In other words, the peak at 2.4 GeV/c 
in fig.10 can be reproduced after excluding all interaction 
terms but the quoted two. 
For $k_T$ $<$ 1.8 (not shown in fig.10) the asymmetry is never 
larger than about 0.2. 

6) The fall of the asymmetry at small $k_T$ is an unavoidable 
consequence of the reduced size of the SSA in elastic scattering. 
For $k_T$ $>$ 3 GeV/c it is a consequence of the lack of hard 
interfering contributions, extracted from elastic scattering. 
It must be remarked, however, that in the case of large $k_T$ 
this lack is not constrained by elastic scattering data. 
Elastic scattering data do not show any decrease at large $k_T$. 
Simply, we do not know what happens for $k_T$ $>$ 3 GeV/c. 
But in fitting figs. 7 and 8 I have adopted a ``minimal'' 
approach, only introducing the strictly necessary interaction 
terms to reproduce the presently available data ranges. Both 
in the case of elastic and inclusive data this approach may 
lead to underestimation of asymmetries at large $k_T$. 

7) Curves in figs. 8 and 10 present qualitative differences 
at intermediate $k_T$. This is due 
to two facts: 
(i) the presence of a fourier transform $exp(ix\xi)$ 
in the inclusive case, that changes the phase 
properties of the interaction terms, (ii) 
eq.\ref{eq:glauber1d} does not contain 
the ``$-1$'' factor of eq.\ref{eq:glauber1c}, so in the 
inclusive case we 
have interference between no-rescattering and rescattering 
terms. 

8) As specified in the Introduction, it is not possible to 
know whether the predicted effect is a leading twist or a 
higher twist one. So, if one names ``Sivers function'' 
the function $q_S(x,k_T)$ that may be extracted from the 
asymmetry between the two curves of fig.9 using 
eq.\ref{eq:sivers}, the result 
shown in fig.10 may be read as a Sivers function in 
the sense that it $mimics$ a Sivers function at finite 
energies.

\subsection{Convolution with a partner 
distribution} 

In all the relevant phenomenological 
cases, 
the calculated functions must 
be convoluted with a partner function $f(x',\vec k_T')$, 
so that the experimentally detected transverse 
momentum is $\vec Q_T$ $=$ $\vec k_T$ $+$ $\vec k_T'$. 

In a Drell-Yan event 
$f$ is the momentum distribution of an antiquark or of a gluon. 
In the case of meson production 
in hadron-hadron scattering, 
$f$ will be a convolution of distribution and fragmentation  
effects, so that $x'$ and $\vec k_T'$ will be product/sum 
of distribution and fragmentation variables. In both cases, we 
must distinguish the cases where $k_T'$ derives from 
Fermi motion ($k_T'$ $\approx$ 0-0.7 GeV/c) or from hard gluon 
secondary radiation ($k_T'$ $>>$ 0.5 GeV/c). 

The former one  
will be the case for $Q_T$ $<$ 3 GeV/c and 
for beam 
energies $<<$ 100 GeV. 
It is then realistic to imagine a gaussian 
distribution with pure Fermi motion origin like 
the following one: 
\begin{equation}
f(x',\vec k_T')\ \propto\ e^{-(k_T/k_0)^2}\label{eq:smearing1}.
\end{equation}
So, fig.11 in the following refers to this situation.  
For larger $Q_T$ and larger beam energies 
the requirement of a transverse momentum with 
dominating Fermi motion origin is not justified.

\begin{figure}[ht]
\centering
\includegraphics[width=9cm]{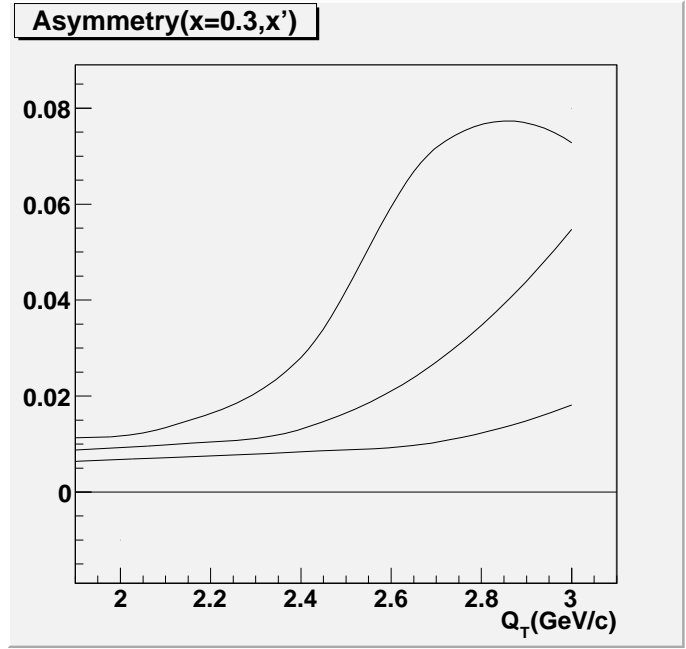}
\caption{Estimate of a possible phenomenological asymmetry, 
e.g. in Drell-Yan or in meson production. 
Here the two quark distributions reported in fig.9 have 
been convoluted with a function $f(x',\vec k_T')$ $=$ 
$exp[-(k_T'/k_o)^2]f(x')$ (see text), and $\vec Q_T$ $=$ 
$\vec k_T$ $+$ $\vec k_T'$. The three curves correspond 
to $k_o$ $=$ 0.4, 0.5, 0.6 GeV/c (the asymmetry decreases for 
increasing $k_o$). 
The partner function $f(x',\vec k_T')$ 
may be the distribution function for an antiquark or gluon, or 
a combination of distribution and fragmentation effects, or 
also include recoil effects by radiation of undetected gluons. As evident, 
the asymmetry is visible if the momentum spread has Fermi motion origin 
($k_o$ $<<$ 1 GeV/c) but not in the case where the momentum 
spread derives from gluon radiation ($k_o$ $>$ 1 GeV/c). 
\label{phenomenology}}
\end{figure}

To correctly calculate a convolution for 
assigned $x$ and $x'$, I need the 
values of $q(x,\vec k_T)$ (the function appearing in 
eq.\ref{eq:sivers} and plotted in fig.9) over a wide 2-dimensional 
range of $\vec k_T$. 
As previously observed, the numerical calculation of 
$q(x,\vec k_T)$ is restricted by computational problems, 
that increase at increasing $k_x$ and $k_y$. For 
$x$ $=$ 0.3 and $k_y$ $=$ 0, I have a reliable set 
of values of $q(x,\vec k_T)$ up to $k_x$ $=$ 2.6-2.8 
GeV/c. To estimate a convolution, I make the simplifying 
hypothesis 
that $q(x,\pm\vec k_T)$ depends on $k_y$ 
via a factorized term 

\begin{equation}
q(x,\vec k_T)\ =\ q(x,k_x,0)q'(k_y)\label{eq:smearing2}.
\end{equation}

With this hypothesis,  
the terms that depend on $Q_x$ 
and $Q_y$ separate in the convolution 
$\int dk_xdk_y q(x,k_x,k_y)f(x',Q_x-k_x,Q_y-k_y)$. 
For $Q_y$ $=$ 0, I simply obtain 

\begin{equation}
\int dk_xdk_y \Big[q(x,k_x,k_y)f(x',Q_x-k_x,Q_y-k_y)\Big]_{Q_y=0} 
\ =
\nonumber
\end{equation}
\begin{equation}
=\ const \cdot \int dk_x q(x,k_x,0)f(x',0)e^{-[(Q_x-k_x)/k_0]^2}
\label{eq:smearing3}
\end{equation}

As for the calculation of fig.10, the above integral may be 
calculated for positive and negative $Q_x$. The corresponding 
asymmetry $(G_+-G_-)/(G_++G_-)$ is shown in fig.11. 

Obviously the asymmetry shown 
in fig.11 inherits the numerical uncertainties of the 
calculation of $q(x,\vec k_T)$, and for this reason 
it 
cannot be considered 100 \% reliable at the largest 
reported $Q_T$. 
As for fig.10, the  qualitative 
origin of the asymmetry peak is 
the interference between semi-hard and hard interaction 
terms. The semi-hard term decreases for $k_T$ $>$ 
2.5 GeV/c, and so indirectly for $Q_T$ $>$ 3 geV/c. 
So, qualitatively fig.11 shows nothing strange, but 
it cannot be considered precise 
for $Q_T$ $>$ 2.7 GeV/c. 

Comparing fig.10 with fig.11 one notices that in the latter 
case we have a predictable  
spread of the asymmetry peak 
towards large $Q_T$, but 
not towards small $Q_T$. This is due to the steep 
decrease of the quark distribution $q(x,k_x,0)$ (see fig.9) for 
$k_x$ $=$ 2$-$2.5 GeV/c. Events with $k_x$ $=$ 
2.5 GeV/c and $k_x'$ $=$ $-$0.5 GeV/c  
are much less frequent than events 
with $k_x$ $=$ 2 GeV and $k_x'$ $=$ 0, so the former kind of 
events does not influence the region $Q_x$ 
$\approx$ 2 GeV/c. 

For this reason, if the average $k_T'-$spread of the partner 
distribution 
$f(x',\vec k_T')$ is increased from 0.4 GeV/c to larger 
values, the peak asymmetry decreases. So, the phenomenological 
effects are expected to depend in a marked way on the specific 
measurement.\footnote{In 
the meson production case  
things are also complicated by 
the presence of direct contributions to the asymmetry  
from the fragmentation 
side\cite{CollinsFunction}.} 

Referring to the discussion in point (6) of subsection 5.1, 
it must be remarked that the fact that the found asymmetries 
decrease for $Q_T$ $<$ 2 GeV/c is a necessary consequence 
of the small (average) values of SSA in elastic scattering 
for $Q_T $ $<$ 1.8 GeV/c (fig.8). So, in the case of a 
relevant asymmetry (over 5 \%) measured at $Q_T$ $<$ 
1 GeV/c, I would exclude that it may have the 
origin that is described here. 

On the contrary, there are no 
experimental constraints on the asymmetry values from 
elastic data at $Q_T$ $>$ 3 GeV/c, where more interaction 
terms could be present, aside of those considered here. 
I have adopted 
a ``minimal'' approach, in the sense of introducing just those 
interaction terms that are strictly necessary for reproducing 
the visible data. 
So the prediction reported here could  
underestimate asymmetries on the right side of the 
peak in figures 10 and 11. 

\subsection{Discussion} 

As previously observed, I do not imagine the term calculated 
by me to saturate the Sivers function. Rather, it is a 
contribution to it in a well-defined kinematic region. 
For the following discussion, I will name ``AB-term'' this 
contribution. 

I will discuss the Drell-Yan application, 
since in this case we have some proved factorization 
statements\cite{CollinsSoperSterman,Bodwin}, a simple 
connection with the 
SIDIS (leptoproduction) case\cite{Collins02}, absence of 
final state effects, a reference hard 
scale $Q$ (the dilepton mass) 
for $Q_T$, and a connection between the large-$Q_T$ 
and the 
small-$Q_T$ behavior of the Sivers asymmetry\cite{JQVY06}. 

From figs. 10 and 11 it is evident that the AB-term may 
be relevant in the region $\Lambda$ $<<$ $Q_T$ $<<$ $Q$, 
for reasonable values of $Q$. $\Lambda$ may be any 
soft scale parameter. 
This region has been studied\cite{JQVY06} because here the 
twist-3 model by refs.\cite{EfremovTeryaev82,QiuSterman91} 
and the Sivers function scheme have 
overlapping regions of consistency. I name "semihard 
$Q_T$ region" the above region. 

The previous calculations refer to proton-proton collisions. 
As far as it can be considered scale-independent, the 
AB-term should be present in SIDIS too according with the 
change-of-sign rule stated in \cite{Collins02} for the 
leading-twist Sivers function. 

In Drell-Yan we have two relevant scales: $Q$ and $Q_T$. 
It is normally admitted\cite{Sivers} that the Sivers asymmetry 
is 
power-suppressed in $Q_T$, but the associated Sivers function 
is anyway quoted as "leading twist" if it is not 
power-suppressed with respect to 
$1/Q$ for fixed $x$ and $k_T$. 
Since $Q$ is related to the 
squared c.m. energy $s$ via the 
scaling variables $x_1$ and $x_2$ ($Q^2$ $=$ $x_1x_2s)$) 
``leading twist''  
means that in eq.\ref{eq:sivers} $q_S$ 
depends on $s$ at most logarithmically (for 
fixed $x$ and $k_T$). 

In this work $s$ is hidden in eq.\ref{eq:glauber1bis}, 
in the dependence of 
$T'$ on the momentum $k_o$ of the projectile quark 
(in a frame where the target hadron is at rest) 
This dependence means 
that the 
strength parameters $\delta_n$ and $\alpha_n$ in eqs. \ref{eq:delta_n} 
and \ref{eq:alpha_n}
are 
in principle functions of $s$. These parameters are extracted 
from the data in fig.7 and 8, so they can be considered as 
$s$-independent if these data do not depend on the beam energy. 
The data reported in fig.7 are stable for beam energy 20-50 GeV, 
and would show logarithmic changes at larger energies. 
Those of fig.8 are stable in the range 10-30 GeV, but we 
have no similar data at larger energies. So, I cannot 
presently 
establish whether the strength of the spin-orbit terms is 
$s$-independent or is not. If it is, the function shown in 
fig.10, with a change of sign, is a contribution to the 
Sivers function in leptoproduction. 

If I decided to calculate directly (with a technique that is 
reasonably similar to the one adopted here) the SIDIS 
$\pi^+$ production on proton, 
this would be 
the crossed process of $\pi^-$-proton Drell-Yan. For the 
Sivers asymmetry in these two processes the calculation 
performed within my scheme would respect Collins' rule 
(since the only change between the two cases would be in 
the integration path for the eikonal factor). 
However, the calculation of any of these two processes 
would require additional assumptions, since the elastic 
data of figs. 7 and 8 only constraint proton-proton ISI.

An obvious problem with figs. 10, 11 is the non regular 
rise of the asymmetry with respect to $Q_T$. 
A nonzero AB-term in the semihard region does not exclude the 
simultaneous presence of a Sivers function like those 
proposed in refs.\cite{BoerBrodskyHwang03,GambergGoldsteinOganessyan03,
BacchettaSchaeferYang04,LuMa04,Yuan03,Courtoy08}
filling the soft $Q_T$ region where the AB-term is small. 
Indeed, we may have mechanisms that are able to produce a 
single spin asymmetry in inclusive scattering but become 
ineffective, or effective but scarcely visible, when applied 
to elastic scattering. These mechanisms would escape the 
presented analysis. 

On the other side, although nothing in my model forbids a 
nonzero AB-term at small $Q_T$, it is a 
matter of fact that, whatever mechanism produces a nonzero 
analyzing power in elastic scattering (data in fig.8), this 
mechanism has its top relevance in the region $Q_T$ $>$ 2 GeV/c, 
and rather small effects at $Q_T$ $<$ 1 GeV/c. So, to invent a 
model that transforms these small effects, observed in elastic 
scattering at small $Q_T$, into relevant effects in inclusive 
processes at the same $Q_T$ would not be trivial. 

A over-simplified way to reproduce the physics described 
in this paper 
can be to imagine that, before the hard inclusive event, 
the two colliding protons scatter elastically and remain 
almost on shell up to the hard 
event.\footnote{This process is power-suppressed in $Q_T$ 
with respect to the more general process considered in 
this paper, since 
it requires re-formation of the proton bound state after 
ISI and before the e.m. hard 
scattering.} 
We know from 
the data of fig.8 that after the elastic scattering 
the space distribution of the colliding protons 
depends on the normal spin of the initial state. This is 
the entrance way to a nonzero Sivers asymmetry, since these 
protons enter the later hard scattering with asymmetric 
momentum distribution. The same data in fig.8 tell us that 
this effect is remarkable only when the $Q_T$ that is exchanged 
in the elastic scattering is semihard. We note that the 
effect is present also if quarks are completely collinear 
in the initial state, since what matters is the $Q_T$ 
exchanged in ISI. 
So one has a physical picture where the typical event 
characterized by a nonzero AB-term has small ``primordial'' 
$Q_T$, and a semihard $Q_T$ produced in ISI. 

For the models of the Sivers asymmetry in Drell-Yan 
that are present in my reference list, we may 
distinguish two classes: 

(A) models where the transverse momentum dependence of 
the cross section is entirely of Fermi motion, ``primordial'', 
origin. Here $Q_T$ $=$ $k_{T1}$ $+$ $k_{T1}$ 
is the sum of the transverse 
momenta intrinsically associated with the distribution 
functions of the colliding quark and antiquark. 

(B) models where the distribution functions are initially 
collinear, the $Q_T$-dependence of the cross section 
is associated with the hard nucleon-nucleon 
interaction (e.g. gluon radiation accompanying the hard 
vertex) and some arguments allow one to stick (part of) 
this effect to the individual quark distributions. 

All the models extending the one by \cite{BrodskyHwangSchmidt02} 
belong to group A. The calculation in \cite{JQVY06} belongs to 
group B. 
The model discussed here 
is mid-way between the two groups: the unperturbed 
quark distribution is narrow, but not fully collinear (it 
has a gaussian tail and transverse range 1 fm), and the event 
numbers at semihard $Q_T$ are strongly enhanced by ISI. The 
Glauber-Gribov approximation allows me to stick this 
enhancement to the quark distributions. 
In absence of ISI the curves in fig.9 would be lower in 
magnitude for $K_T$ over 1 GeV/c.

\section{Conclusions} 

Summarizing, starting from the assumption that the quark 
total angular momentum is 
dominantly oriented as the parent hadron spin, it is possible to 
build a nonzero Sivers-like asymmetry via 
mean field initial state interactions of scalar and 
spin-orbit kind. 

The specific form of the interactions, together with the 
values of the parameters, is here extracted from the phenomenology 
of proton-proton elastic scattering at 20-50 GeV. 
When these interactions 
are applied to transverse momentum dependent quark distributions, 
they produce an asymmetry that is important at transverse momenta 
2-3 GeV/c, and reduces to very small values for  
smaller transverse momenta. 
It is not 
possible to estimate what happens over 3 GeV/c, for the lack of 
corresponding data in elastic proton-proton scattering. 

With the present day knowledge of spin-orbit hadronic interactions, 
it is not possible to know whether their effects persist at 
very large energies. Consequently it is not possible to establish 
whether a Sivers-like asymmetry generated by them is a leading 
twist one or just an intermediate energy effect. 

\section{Appendix A}

Solutions of the Dirac equation have in general 
four independent components. This is related with the existence of 
two solutions of the scalar equation $E^2 = p^2 + m^2$ and 
of two independent helicity states for each energy. 
In the two limiting cases 

\begin{equation}
E\ \approx\ m\ +\ {p^2 \over {2m}},
\end{equation}
\begin{equation}
E\ \approx\ p_z\ +\ {m_T^2 \over {2p_z}},
\end{equation}

\noindent
the structure simplifies. Although the use of a standard 4-component 
Dirac formalism may allow one to exploit a large set of well-known 
relations for trace calculation etc, it hides the fact that two 
components only are independent in the ultrarelativistic or 
nonrelativistic spinors. 

Writing a 4-spinor in standard representation as 
\begin{equation}
\Psi\ \equiv\ (\phi,\chi)
\end{equation}
the two 2-spinors satisfy the coupled equations 
\begin{eqnarray}
\begin{array}{cc}
(E - m) \phi\ =\ \vec p\cdot\ \vec \sigma \chi \\
(E + m) \chi\ =\ \vec p\cdot\ \vec \sigma \phi \\
\end{array}
\label{eq:dirac}
\end{eqnarray}
equivalent to the Dirac equation. 

We have the known nonrelativistic limit $\chi$ $\rightarrow$ 
0 for particles, and the opposite limit for antiparticles. This 
decouples each other 
spin states, since the second one of eqs.\ref{eq:dirac} loses relevance, 
and in the first one the right-hand side may be neglected. So, only 
interactions may remix spin states. 

As well known, 
in the opposite limit $E$ $>>$ $m$ we have two relevant states 
$\phi\pm\chi$ that decouple each other. The nature of these 
states is clear after 
introducing helicity $\hat n\cdot \vec \sigma$. The u.r. limit 
may be reached for $E$ $\approx$ $\pm p_z$. For $E$ $\approx$ 
$+p_z$, $\phi+\chi$ coincides with one of the two helicity 
states, and $\phi-\chi$ with the other one. For $E$ $\approx$ 
$-p_z$, the correspondence is the opposite. 

Instead of using helicity basis, 
I write the above equations \ref{eq:dirac} in terms of 
eigenstates of the spin projection along the 
$x-$axis\footnote{ 
In this part of the work I project the transverse spin along the $x-$axis 
and the transverse momentum along the $y-$axis. This simplifies 
intermediate passages if one wants to reproduce oneself 
the above relations. In the rest of the work I project the spin 
along the $y-$axis to conform to ordinary 
treatments. 
}
$\phi_{x\pm}$ $\equiv$ $\phi_{T\pm}$:
\begin{equation}
\phi_{T-} \ =\ (1,1)/\sqrt{2},\hspace{0.5truecm} 
\phi_{T-} \ =\ (1,-1)/\sqrt{2} 
\end{equation}
These 2-spinors 
satisfy the overlap relations 
\begin{eqnarray}
\begin{array}{cc}
\sigma_z \phi_{T+}\ =\ \phi_{T-},\hspace{0.5truecm} 
\sigma_z \phi_{T-}\ =\ \phi_{T+} \\
\sigma_y \phi_{T+}\ =\ -i\phi_{T-},\hspace{0.5truecm} 
\sigma_y \phi_{T-}\ =\ i\phi_{T+}
\end{array}
\label{eq:rotate}
\end{eqnarray}
and the same for the corresponding $\chi_{T\pm}$ terms. 

In the u.r. limit, from eq.\ref{eq:dirac} one gets the relations 
\begin{equation}
\phi_{T+}\ \approx\ \chi_{T-}, 
\hspace{0.5truecm}  \phi_{T-}\ \approx\ \chi_{T+}
\end{equation}
meaning that a spinor polarized along the transverse axis has 
the approximate form 

\begin{equation}
\Psi_{T+}\ \equiv\ (\phi_{T+},\chi_{T+}),\ 
\approx\ (\phi_{T+},\phi_{T-}),\hspace{0.5truecm} 
\Psi_{T-}\ \approx\ (\phi_{T-},\phi_{T+})
\end{equation}

\noindent
More precisely, in presence of a nonzero transverse momentum 
$p_y$ $<<$ $p_z$ that is also orthogonal to the normal 
spin, eqs.\ref{eq:dirac} become 
(exploiting the previous eqs.\ref{eq:rotate}): 

\begin{eqnarray}
\begin{array}{cc}
E\phi_{T+} - p_z \chi_{T-}\ +\ m\phi_{T+} +i p_y\chi_{T-}\ =\ 0, \\
E\chi_{T+} - p_z \chi_{T-}\ +\ m\phi_{T+} +i p_y\chi_{T-}\ =\ 0, 
\end{array}
\label{eq:dirac2}
\end{eqnarray}

\noindent
so that $\phi_{T+}$ and $\chi_{T-}$ decouple from 
$\phi_{T-}$ and $\chi_{T+}$ (remark: 
$\phi_{T+}$ and $\chi_{T-}$ are pieces of different 4-spinors). 
Writing $E$ $=$ $(E - p_z)$ $+$ $p_z$, the former of the previous 
two equations becomes 

\begin{equation}
p_z(\phi_{T+} - \chi_{T-})\ +\ (E-p_z)\phi_{T+}\  
+\ m\phi_{T+} +i p_y\chi_{T-}\ =\ 0.
\end{equation}

At $O(1/E)$ this confirms the equality between $\phi_{T+}$ and  
$\chi_{T-}$. Substituting $\chi_{T-}$ $=$ $\phi_{T+}$, 
and exploiting $E-p_z$ $\approx$ $m_T^2/2p$ $=$ 
$(m+ip_y)(m-ip_y)/2p$ one may also quantify 
the difference between the two spinors, that is 
$O(\phi_{T+} p_y/p_z)$. 

For the $\gamma_+$ projection we have 
(proportionality factors substitute 
powers of $1/\sqrt{2}$)
\begin{eqnarray}
\Psi_{T+}^+\gamma_0\gamma_+\Psi_{T+}\ \propto\ 
\Psi_{T+}^+
\left(
\begin{array}{cc}
1 &  \sigma_z \\
\sigma_z &  1 
\end{array}
\right)
\Psi_{T+}\ =
\end{eqnarray}
\begin{equation}
\propto\ 
|\phi_{T+}|^2\ +\ 
\phi_{T+}^+\chi_{T-}\ +\ 
\chi_{T+}^+\phi_{T-}\ +\ 
|\chi_{T+}|^2\ 
\propto\
|\phi_{T+}|^2\ +\ |\phi_{T-}|^2.  
\end{equation}
and the same for $\Psi_{T-}$. 
In addition, 
\begin{equation}
\Psi_{T+}^+\gamma_0\gamma_+\Psi_{T-}\ =\  
\Psi_{T-}^+\gamma_0\gamma_+\Psi_{T+}\ =\  0.
\end{equation}

So, if a generic 4-spinor $\Psi$ has the form 
\begin{equation}
\Psi\ \equiv\ \psi_+ \Psi_{T+}\ +\ \psi_- \Psi_{T-}
\end{equation}
we have, within corrections $O(1/E)$,  
\begin{equation}
\bar\Psi\gamma_+\Psi\ \approx\  
\Psi^+\gamma_0\gamma_+\Psi\ \approx\  
|\psi_+|^2\ +\ |\psi_-|^2,
\label{eq:trace10}
\end{equation}
i.e. the result exploited in eq.\ref{eq:sivers3} of the present work. 

\section{Appendix B}

The exploitation of the Glauber technique in DWBA 
is better suited in nonrelativistic or 
ultrarelativistic form. In both limits it is 
often possible to 
approximate the kinetic energy in such a way that (i) 
a large part of the boost energy is removed 
from the problem, 
(ii) interactions may be inserted 
as additive contributions, to the zero or to the plus 
component of the energy-momentum vector.  
When (i) is true, the two treatments are equally 
effective. 
When it is not so, corrections are needed in both 
cases\cite{Gribov69}. 

In the n.r. case, one normally uses a 
linearized Schroedinger equation, but relativistic 
kinematics. Since one always works with momenta, 
and interactions are directly fitted within this scheme, 
the hamiltonian is never involved and 
it is not surprising that this scheme may work  
in relativistic conditions. A detailed discussion of this  
and related references may be found in \cite{BREikonal}. 

Limiting the discussion here to the scalar case for simplicity, 
in the n.r. case one of the possible starting points for the 
Glauber treatment 
is a linearized form of the Schroedinger equation, via 
the 
approximation  
\begin{equation} 
(k^2 - \hat p^2)\psi\ \approx\ 2 k (k - \hat p_z)\psi
\end{equation}
where the kinetic energy is $(\hbar k)^2/2m$. Rescaling the 
potential energy as $2kV$ $\equiv$ $2mU/\hbar^2$ the linearized 
Schroedinger equation 
is
\begin{equation} 
\hat p_z \psi\ =\ k (1+V) \psi, \hspace{0.5truecm}
V\ <<\ 1. 
\end{equation}
with 
stationary solution (for $V$ $<<$ 1) 
\begin{equation} 
\psi^{in}\ =\ exp\bigg( i k z + i \int_{-\infty}^z V(b,z') dz'\bigg),
\end{equation}
that 
is the n.r. version of the wavefunction $\psi_o$ appearing in 
eq.\ref{eq:glauber1b}. The corresponding solution $\psi^{out}$
has $\int_{-\infty}^z$ substituted by $\int_z^{-\infty}$. 

For the calculation of the matrix element 
$<\psi_f|\psi_o>$ in eq.\ref{eq:glauber1b}, the eikonal direction $\hat z$ 
of the dominating $exp(ikz)$ factor 
is rotated to $k\hat z \pm (q/2) \hat x$ in the initial or final 
wave, while the eikonal distortion factor is still approximately 
calculated along the direction $\hat z$ that is the average 
direction between the initial and final wave vectors. 

For the insertion of the interactions in the 
u.r. case, there are different possibilities. The one adopted 
here is to insert the interaction $T$ into the u.r. 
equation
\begin{equation} 
(x P_+\ -\ \hat p_+)\psi\ \approx\ 0,  
\end{equation}
that 
becomes
\begin{equation} 
(\hat p_+\ -\ xP_+\ +\ P_+T) \psi\ =\ 0. 
\end{equation}
When 
this equation is written in terms of the rescaled variable $\xi$ 
$=$ $P_+z_-$ we get the 
solution
\begin{equation} 
\psi\ =\ exp\bigg( -i x \xi + i \int_{-\infty}^\xi T(b,\xi') d\xi'\bigg)
\end{equation}
appearing 
in eq.\ref{eq:glauber1b} 
(rotated $k \hat z$ $\rightarrow$ $k\hat z \pm (q/2) \hat x$). 

Evidently, the previous u.r. and n.r. solutions are equivalent, once 
the interaction factors $V$ and $T$ are correctly related each 
other. 

The tricky point\cite{Gribov69} is that in both cases one 
assumes that some observable, associated with the longitudinal 
highly relativistic motion, is $constant$. 
In the n.r. case it is $k$, in the u.r. case, it is $x$. 
``Constant'' means ``conserved  
during all the development of the motion''. This way, the corresponding 
operator is substituted by its eigenvalue in the wave equation. 

Multiple interactions may initially  
convert the initial quark state with mass $m$ 
into a more complex intermediate state with invariant mass $M$ 
(e.g., a quark plus a quark-antiquark pair). If this state 
is converted again into a single quark state, it contributes 
to elastic scattering. 
In highly relativistic conditions we have changes 
of $p_z$ (in intermediate states) 
\begin{equation}
\Delta p_z\ \sim\ {{M^2-m^2} \over p_z}, \hspace{0.5truecm} 
M\ <\ M_{up}
\end{equation}

\noindent
where $M_{up}$ is a reasonable upper bound for the 
spectrum of intermediate states that may play some role. 

The classical Glauber regime requires $M_{up}$ $<<$ $p_z$. 
In this case $p_z$ $\approx$ constant in the interactions, 
the $O(p_z)$ term has no dynamical role, 
and the really relativistic part of the quark wavefunction 
disappears from the problem. 

The Gribov regime is the one where in the limit 
$p_z$ $\rightarrow$ $\infty$,  
$M_{up}$ $=$ 
$\alpha p_z$ with small but nonzero $\alpha$. 
Then we have finite $O(\alpha)$ changes in $x$ during the intermediate 
stages of the process, invalidating the above assumption that 
$x$ is a constant of motion. 

Since a full 
inclusion of inelastic channels is a difficult and unsolved 
problem, such channels 
are normally taken into account via two techniques: 
the first one is to introduce a discrete and limited set of  
intermediate non-elastic channels\cite{KL78}. The other one is to 
rewrite the interaction terms in a non-hermitean 
form. 

The latter treatment is $formally$ 
equivalent to a multi-channel treatment, 
as far as one is interested in the 
evolution of the elastic channel only. For nonrelativistic nuclear 
physics, this was demonstrated by Feshbach\cite{Feshbach}. 
The scheme employed by him is quite abstract, and 
it can be generalized practically to any 
algebraic set of coupled equations describing the evolution 
of a state vector. So this equivalence may be considered general. 
The drawback of this generality is 
that it is practically impossible to use Feshbach's arguments 
to calculate the 
form of the effective interaction from first principles. 

So, in this work I had to guess a general form for the 
interaction operator $\hat T$, and fit it on elastic proton-proton 
scattering according to eq.\ref{eq:glauber1c}. 

A last relevant point is that the treatment via the single channel 
effective 
potential does not transport the kinematics of the rescattering 
process out of the 
Glauber region (the price for this is that momenta are complex).
Rescattering in the Glauber region does not create problems with 
factorization, 
at least in the Drell-Yan 
case for which an exhaustive discussion of the 
problem exists\cite{CollinsSoperSterman,Bodwin}.



\end{document}